\documentclass[review,onefignum,onetabnum]{siamonline190516}

\usepackage{lipsum}
\usepackage{amsfonts}
\usepackage{graphicx}
\usepackage{epstopdf}
\usepackage{mathtools}
\usepackage{bm}
\usepackage{amsopn}
\usepackage{algorithmic}
\ifpdf
\DeclareGraphicsExtensions{.eps,.pdf,.png,.jpg}
\else
\DeclareGraphicsExtensions{.eps}
\fi

\usepackage{enumitem}
\setlist[enumerate]{leftmargin=.5in}
\setlist[itemize]{leftmargin=.5in}

\DeclareMathOperator*{\argmin}{arg\,min}
\DeclareMathOperator*{\argmax}{arg\,max}
\DeclareMathOperator*{\solve}{solve}

\newsiamremark{remark}{Remark}
\newsiamremark{hypothesis}{Hypothesis}
\crefname{hypothesis}{Hypothesis}{Hypotheses}
\newsiamthm{claim}{Claim}

\headers{Certified Dimension Reduction
	for Bayesian Updating with the
	Cross-Entropy Method}{Max Ehre, Rafael Flock, Martin Fu{\ss}eder, Iason Papaioannou, Daniel Straub}

\title{Certified Dimension Reduction
	for Bayesian Updating with the
	Cross-Entropy Method\thanks{Submitted to the editors March 12, 2022.
		\funding{We acknowledge support by the German Research Foundation (DFG) through Grants STR 1140/11-1 and PA 2901/1-1.}}}
\author{Max Ehre\thanks{Technical University of Munich, School of Engineering and Design, Engineering Risk Analysis Group
		(\email{max.ehre@tum.de}, \email{iason.papaioannou@tum.de}, \email{straub@tum.de}).}
	\and Rafael Flock\thanks{Technical University of Denmark, Department of Applied Mathematics and Computer Science, DTU Compute
	    (\email{raff@dtu.dk}).} 
    \and Martin Fu{\ss}eder\thanks{Technical University of Munich, School of Engineering and Design, Chair of Structural Analysis
    	(\email{martin.fusseder@tum.de}).} 
	    \and Iason Papaioannou\footnotemark[2]
	\and Daniel Straub\footnotemark[2]}

\crefname{equation}{Eq.}{Eqs.}
\crefname{figure}{Fig.}{Figs.}
\crefname{tabular}{Tab.}{Tabs.}
\crefname{section}{Section}{Sections}
\crefname{subsection}{Subsection}{Subsections}
\crefname{algorithm}{Alg.}{Algs.}
\crefname{chapter}{Chapter}{Chapters}

\newcommand{\proj}{\mathbf{P}_{r}}
\newcommand{\projperp}{\mathbf{P}_{\perp}}
\newcommand{\tran}{\mathrm{T}}

\definecolor{subLIS}{rgb}{0.1216    0.3137    0.7020}
\definecolor{subCS}{rgb}{0.0314    0.3961    0.0392}

\begin{document}
\nolinenumbers
\maketitle

\begin{abstract}
In inverse problems, the parameters of a model are estimated based on observations of the model response.  The Bayesian approach is powerful for solving such problems; one formulates a prior distribution for the parameter state that is updated with the observations to compute the posterior parameter distribution. Solving for the posterior distribution can be challenging when, e.g., prior and posterior significantly differ from one another and/or the parameter space is high-dimensional. We use a sequence of importance sampling measures that arise by tempering the likelihood to approach inverse problems exhibiting a significant distance between prior and posterior. Each importance sampling measure is identified by cross-entropy minimization as proposed in the context of Bayesian inverse problems in Engel et al. (2021). To efficiently address problems with high-dimensional parameter spaces we set up the minimization procedure in a low-dimensional subspace of the original parameter space. The principal idea is to analyse the spectrum of the second-moment matrix of the gradient of the log-likelihood function to identify a suitable subspace. Following Zahm et al. (2021), an upper bound on the Kullback-Leibler-divergence between full-dimensional and subspace posterior is provided, which can be utilized to determine the effective dimension of the inverse problem corresponding to a prescribed approximation error bound. 
We suggest heuristic criteria for optimally selecting the number of model and model gradient evaluations in each iteration of the importance sampling sequence. We investigate the performance of this approach using examples from engineering mechanics set in various parameter space dimensions.
\end{abstract}

\begin{keywords}
  Bayesian inverse problems, high dimensions, cross-entropy method, importance sampling, certified dimension reduction
\end{keywords}
%
\begin{AMS}
  62F15, 62L12, 62P30, 60G60, 65C05
\end{AMS}

\graphicspath{{figures/}{tikz/}{plots/}}

\section{Introduction}
\label{sec:intro}
We consider inverse problems in the context of a computational model $f$ with $\bm{y} = f(\bm{\theta})$. That is, we want to characterise a cause (parameters of the computational model, $\bm{\theta}$) based on observations of the corresponding effects or consequences of said cause (output of the computational model $\bm{y}$). 
An example is a structural system represented with a finite element model that is parameterized by loads, geometric and material properties $\bm{\theta}  \in \mathcal{X} \subseteq \mathbb{R}^d$ and that produces outputs such as stresses and deflections $\bm{y} \in \mathcal{Y} \subseteq \mathbb{R}^m$.
In the majority of applications, we cannot expect the inverse problem to be well defined, i.e., there need not be a solution, the solution may not be unique or it might be very sensitive to the observations \cite{Stuart2010}. To further complicate the matter, in practice, observations are often incomplete and/or contaminated with noise. 
Here, we focus on the Bayesian approach to inverse problems, which offers a consistent framework for incorporating  both noisy and incomplete observations as well as addressing ill-posedness by regularizing the problem using prior information \cite{Kaipio2005,Stuart2010}.
\cite{Kaipio2005,Stuart2010} discuss the Bayesian inverse problem (BIP) in infinite-dimensional settings, while in practice, we usually retreat to the finite-dimensional case by means of discretizing infinite-dimensional random objects such as random fields and processes. Hence, we focus on finite-dimensional BIPs in this work.
\\~\\
We represent $\bm{\theta}$ and $\bm{y}$ as real-valued random vectors $\bm{\Theta}:\mathcal{X},\mathcal{B}(\mathcal{X}) \to \mathbb{R}$ and $\bm{Y}:\mathcal{Y},\mathcal{B}(\mathcal{Y}) \to \mathbb{R}$, where $\mathcal{B}(\cdot)$ is the Borel $\sigma$-algebra,
and we assume the probability measures $\mathbb{P}_{\bm{\Theta}}$, $\mathbb{P}_{\bm{Y}}$ to be absolutely continuous with respect to the respective Lebesgue measures on $\mathcal{B}(\mathcal{X})$ and $\mathcal{B}(\mathcal{Y})$. We then may use the associated probability density functions (PDF) $p(\bm{\theta})$ and $p(\bm{y})$ to characterize $\bm{\Theta}$ and $\bm{Y}$.
\\~\\
We start by placing a prior distribution on $\bm{\theta}$ by defining the prior PDF $p_0(\bm{\theta}): \mathbb{R}^d \rightarrow \mathbb{R}_{\geq 0}$. 
As the name suggests, $p_0(\bm{\theta})$ formalizes any information one may have on $\bm{\theta}$ \emph{prior} to considering any observations. 
This information may come as the outcome of an expert elicitation, selection rules \cite{Ohagan2019} and/or guiding principles to construct noninformative priors such as Jeffreys's priors \cite{Jeffreys1961} or priors satisfying the maximum entropy principle \cite{Jaynes1982}. While many of these principles rest on the idea to minimize the influence the prior exerts on the posterior distribution and thus aim at `letting the data speak', it is flat/weak priors in particular that can lead to overly confident inference results \cite{Gelman2020}. A single layer of priors may not do justice to complex models with a large number of unobserved variables, in which case \emph{hierarchical models} with several layers of prior distributions can be utilized \cite[Section 2.8]{Gelman2013}.
\\~\\
Next, one or several observations of $\bm{y}$ that we refer to as $\tilde{\bm{y}}$, are represented by the \emph{likelihood} $L(\bm{\theta}) \vcentcolon  = p(\tilde{\bm{y}} | \bm{\theta}): \mathbb{R}^d \rightarrow \mathbb{R}_{\geq0}$, which states how likely these observations are to occur under any given set of parameters $\bm{\theta}$. In Bayesian inverse problems, the likelihood will be a function of $f$, thereby facilitating the backpropagation of information on outputs of $f$, $\bm{y}$, to its parameters $\bm{\theta}$.
With this, the posterior PDF of $\bm{\theta}$ conditional on observations $\tilde{\bm{y}}$ follows from Bayes' theorem as
\begin{equation}
\label{eq:bayes}
p_{\bm{y}}(\bm{\theta}) \vcentcolon  = p(\bm{\theta} | \tilde{\bm{y}})  = \frac{p(\tilde{\bm{y}} | \bm{\theta})  p_0(\bm{\theta})}{p(\tilde{\bm{y}})} =  \frac{L(\bm{\theta})  p_0(\bm{\theta})}{Z},
\end{equation}
where
\begin{equation}
Z = \int_{\mathcal{X}} L(\bm{\theta})  p_0(\bm{\theta}) \mathrm{d}\bm{\theta}
\end{equation}
is the marginal likelihood of the data also known as the \emph{model evidence}. We assume the evidence is finite $Z < \infty$ and the likelihood is Borel-measurable. 
In the Bayesian approach, solving the inverse problem amounts to computing the posterior distribution of $\bm{\theta}$ and generating samples from $p_{\bm{y}}(\bm{\theta})$. In many instances, the posterior distribution cannot be computed exactly. Instead, sampling approaches such as importance sampling (IS) \cite{Geweke1989} or Markov Chain Monte Carlo (MCMC) \cite{Gilks1995} are used to sample from the posterior and construct estimates of posterior expectations. If prior and posterior distributions differ from one another significantly, constructing efficient \textit{biasing} or \textit{proposal} densities for IS or MCMC, respectively, becomes difficult. Such problems can be addressed by repeatedly applying sampling methods on an artificial sequence of distributions that gradually approach the posterior starting from the prior, namely \emph{sequential Monte Carlo} methods (SMC) \cite{Doucet2001,Neal2001,DelMoral2006}. In \emph{adaptive} SMC \cite{Koutsourelakis2009,Jasra2011,Latz2018}, the distribution sequence is determined during runtime based on intermediate samples.
\\~\\
In SMC approaches, the distributions appearing in the sequence are characterized by samples that are obtained through resample-move steps; samples from each previous distribution are moved via MCMC sampling to obtain samples from the next distribution. However, MCMC produces dependent samples. Alternatively, in \emph{cross-entropy importance sampling} (CE-IS) \cite{Rubinstein2017}, a sequence of parametrized distributions is defined such that each new distribution in the sequence is identified through solving an optimization (cross-entropy minimization) problem. Estimation of the target distribution is then performed with IS using the final fitted parametric density. Hence, CE-IS avoids MCMC sampling and dependent samples. CE-IS was introduced in the context of rare event estimation in \cite{Rubinstein1997} and was recently applied to solve the BIP in \cite{Engel2021}.
 \\~\\
Both acceptance rate and mixing time --- and hence, computational efficiency --- of many MCMC algorithms deteriorate as the problem dimension $d$ increases \cite{Roberts2001,Mattingly2012}; notable exceptions include the \emph{preconditioned Crank-Nicholson} (pCN) sampler \cite{Beskos2008,Cotter2013} and Hamiltonian MCMC \cite{Neal2011}.
Therefore, different approaches have been proposed to reduce the dimension of the inverse problem by identifying low-dimensional subspaces on which the solution to the original problem may be identified in good approximation. While their existence cannot be guaranteed independent of the inverse problem, low-dimensional subspaces frequently occur in BIPs as a result of $f$ being a smoothing operator applied to the input vector $\bm{\theta}$, e.g., in the form of solutions to a set of partial differential equations (PDEs). 
In \cite{Marzouk2009,Uribe2020b} the problem dimension is reduced by representing the prior with a truncated Karhunen-Lo{\'e}ve-expansion. In the context of linear BIPs, \cite{Flath2011,Spantini2015} construct low-rank approximations of the prior-preconditioned Hessian of the log-likelihood thereby exploiting structure in both prior and likelihood.
The \emph{likelihood-informed subspace} method of \cite{Cui2014} extends this approach to nonlinear BIPs based on a low-rank approximation of the posterior-preconditioned Hessian of the log-likelihood.
\cite{Constantine2016} propose a similar approach in which they identify an \emph{active subspace} of the BIP, i.e., a low-rank approximation of the prior-preconditioned negative log-likelihood gradient.
Building on the idea of likelihood-informed subspaces, \cite{Zahm2021} proposes \emph{certified dimension reduction} for nonlinear BIPs and derives an upper bound on the \emph{Kullback-Leibler-divergence} between reduced and full space posterior in function of the subspace dimension.  
 \\~\\
 While the CE-IS approach to BIPs of \cite{Engel2021}  circumvents MCMC altogether, its performance deteriorates with increasing parameter dimension. This is both due to an increasing degeneracy of the IS weights that are used in the context of CE-IS \cite{Rubinstein2009} as well as the rapidly growing number of parameters in the employed distribution models. 
 For example, in Gaussian models with full covariance structure, the number of parameters is $p = d(d+3)/2$, implying that the number of $f$-evaluations required to obtain an accurate fit scale quadratically with $d$. 
 Solving BIPs with CE-IS is therefore only suitable for low parameter dimension.
 \cite{Uribe2020a} uses CE-IS for estimating rare event probabilities of models with large parameter dimension by applying certified dimension reduction.
  \\~\\
  In this work, we devise a scheme to efficiently solve nonlinear BIPs using CE-IS and certified dimension reduction.
  Our method extends the approaches of \cite{Engel2021,Uribe2020a} to address high-dimensional BIPs. Moreover, we introduce heuristic rules for adaptively selecting the number of model and model gradient evaluations  during the simulation.
  \cref{sec:cebu} recapitulates CE-IS for BIPs following \cite{Engel2021}. \cref{sec:cebured} details the certified dimensionality reduction approach for CE-IS and \cref{sec:method} contains a discussion on methodology, algorithmic details and a summary of the final procedure. In \cref{sec:experiments}, we investigate the efficacy of our method on two  structural engineering examples both featuring large parameter dimensions. Concluding remarks are given in \cref{sec:conclusions}.
\section{Cross-entropy-based importance sampling for Bayesian updating}
\label{sec:cebu}
\subsection{Importance sampling and the cross-entropy method}
In this chapter, we briefly describe the \emph{CE-based IS method for Bayesian updating} (CEBU) proposed in \cite{Engel2021}. 
Importance sampling is a variance reduction method for estimating expectations of a function $G(\bm{\theta})$, $\mathbb{E}_{p}[G(\bm{\Theta})]$ \cite[Chapter 5]{Rubinstein2017}.
Note that we use lowercase letters for deterministic variables. We use uppercase letters for matrices and random variables/vectors with the exception of random samples, which are denoted with lowercase letters yet treated as random variables.
Throughout this work we assume all random vectors to be real-valued, i.e., $\bm{X}:(\mathcal{X},\mathcal{B}(\mathcal{X})) \to \mathbb{R}$, where $(\mathcal{X},\mathcal{B}(\mathcal{X}))$ is a measurable space consisting of the outcome space $\mathcal{X}$ and its associated Borel $\sigma$-algebra  $\mathcal{B}(\cdot)$.
Further, we assume probability measures $\mathbb{P}_{\bm{X}}$ to be absolutely continuous with respect to the respective Lebesgue measures on $\mathcal{B}(\mathcal{X})$ so that we may use the associated PDFs $p(\bm{x})$ to describe $\bm{X}$.
Let $q(\bm{\theta})$ be a PDF on $\mathcal{X}$ such that $q(\bm{\theta}) > 0$ whenever $ p(\bm{\theta}) > 0$ and suppose we only know $\psi(\bm{\theta}) = c p(\bm{\theta})$ pointwise with unknown normalizing constant $c$. Then we can write
\begin{equation}
\label{eq:IS}
\mu \vcentcolon  = \mathbb{E}_{p}[G(\bm{\Theta})] = \frac{1}{c}\mathbb{E}_{q}\left[\frac{G(\bm{\Theta})\psi(\bm{\Theta})}{q(\bm{\Theta})}\right] =  \frac{\mathbb{E}_{q}\left[G(\bm{\Theta})w(\bm{\Theta})\right]}{\mathbb{E}_{q}\left[w(\bm{\Theta})\right]},  ~~~\cite[\mathrm{Chapter~9}]{Owen2013}
\end{equation}
where $q$ is termed the \textit{importance}, \textit{auxiliary}, \textit{instrumental} or \textit{biasing} density and $w(\bm{\theta}) = \psi(\bm{\theta})/q(\bm{\theta})$ is referred to as the \textit{likelihood ratio} or \textit{IS weight}. \cref{e:IS} leads to the \emph{self-normalized IS estimate}
\begin{equation}
\label{eq:is_est}
\widehat{\mu}_{\mathrm{IS},q}
=  \frac{1}{n \widehat{c}}\sum_{k=1}^{n} G(\bm{\theta}_k) w(\bm{\theta}_k) 
,~~~\bm{\theta}_k \stackrel{i.i.d.}{\sim} q(\bm{\theta}),
\end{equation}
where an estimate of the normalizing constant is given as $\widehat{c} =  n^{-1}\sum_{k=1}^{n} w(\bm{\theta}_k)$.
For many problems $q$ can be chosen such that \cref{eq:is_est} has lower variance $\widehat{\mathbb{V}}[\widehat{\mu}_{\mathrm{IS},q}]$ than the crude Monte Carlo estimate \cite[Chapter 9]{Owen2013}.
\\~\\
In the context of BIPs $p$ is a posterior distribution \smash{$p_{\bm{y}}$} and the normalizing constant $c$ in \cref{eq:IS} is the model evidence $Z$. \smash{$p_{\bm{y}}$} is the optimal IS density to estimate the model evidence as $\mathbb{V}[\widehat{Z}] = 0$ if $q = p_{\bm{y}}$. Since sampling from the posterior is usually difficult, we continue with a discussion of how to get a parametric $q$ close to $p_{\bm{y}}$.
\\~\\
\cite{Rubinstein1997} proposed finding a parametric IS density $q(\bm{\theta},\bm{v})$ with parameters $\bm{v} \in \mathcal{V}$ by minimizing the \emph{Kullback-Leibler divergence} (KLD) between $q(\bm{\theta},\bm{v})$ and an optimal IS density in the context of rare event probability estimation. \cite{Engel2021} builds on this principle to estimate a parametric distribution that is close to the posterior $p_{\bm{y}}$ as follows.
The KLD between the posterior and the parametric density $D_{\mathrm{KL}}(p_{\bm{y}}(\bm{\theta})||q(\bm{\theta},\bm{v}))$ is defined as \cite{Rubinstein2017}
\begin{equation}\label{eq:DKL}
\begin{split}
D_{\mathrm{KL}}(p_{\bm{y}}(\bm{\theta})||q(\bm{\theta},\bm{v}))
&=  \mathbb{E}_{p_{\bm{y}}}\left[\ln\left(\frac{p_{\bm{y}}(\bm{\Theta})}{q(\bm{\Theta},\bm{v})}\right)\right]\\
&= \frac{1}{Z}\mathbb{E}_{p_0}[L(\bm{\Theta}) \ln(p_{\bm{y}}(\bm{\Theta}))] \underbrace{- \frac{1}{Z}\mathbb{E}_{p_0}[L(\bm{\Theta}) \ln(q(\bm{\Theta},\bm{v}))]}_{\mathrm{cross~entropy}~H(p_{\bm{y}},q(\cdot,\bm{v}))}.
\end{split}
\end{equation}
The first summand on the right-hand side of \cref{eq:DKL} is not a function of $\bm{v}$ so that minimizing $D_{\mathrm{KL}}(p_{\bm{y}}(\bm{\theta})||q(\bm{\theta},\bm{v}))$ is equivalent to maximizing the negative cross entropy:
\begin{equation}
\label{eq:CE_IS_1}
\bm{v} = \argmax\limits_{\bm{v} \in \mathcal{V}} \mathbb{E}_{p_0}[L(\bm{\Theta}) \ln(q(\bm{\Theta},\bm{v}))],
\end{equation}
which conveniently does not depend on the unknown $Z$. An approximate solution of this optimization problem based on samples from $p$ reads
\begin{equation}
\label{eq:CE_IS_2}
\widehat{\bm{v}} = \argmax\limits_{\bm{v} \in \mathcal{V}} \frac{1}{n} \sum\limits_{k=1}^{n} L(\bm{\theta}_k)  \ln(q(\bm{\theta}_k,\bm{v})),~~~\bm{\theta}_k \stackrel{i.i.d.}{\sim} p_0(\bm{\theta}).
\end{equation}
The optimization problem in \cref{eq:CE_IS_2} is usually convex, continuous and the objective function is differentiable with respect to $\bm{v}$ such that identifying $\widehat{\bm{v}}$ is straight-forward.
Closed-form solutions of \cref{eq:CE_IS_2} exist in various situations, e.g., if $q(\bm{\theta},\bm{v})$ is any member of the exponential family \cite[Chapter 8]{Rubinstein2017}. 
\cite{Kurtz2013,Geyer2019} use a Gaussian mixture model in order to capture several disconnected failure regions. \cite{Wang2016} and \cite{Papaioannou2019b} use \emph{von Mises-Fisher} and \emph{von Mises-Fisher-Nakagami} (vMFN) mixture models, respectively, to overcome the poor performance of Gaussian models in high-dimensional rare event probability estimation problems. \cite{Engel2021} test the performance of both Gaussian mixture and vMFN mixture models in the context of the CE method for BIPs and show that although the latter has superior performance in certain high-dimensional settings, the former possesses higher flexibility and is thus able to accurately describe complicated posteriors. In all these works, different variants of expectation maximization are used to solve for $\widehat{\bm{v}}$.
\\~\\
How well $q(\bm{\theta},\widehat{\bm{v}})$ approximates $p_{\bm{y}}(\bm{\theta})$ hinges on how well samples from $p$ can inform the objective function about $L(\bm{\theta})$. 
In other words, if prior and likelihood are not close to one another, we cannot expect the solution of \cref{eq:CE_IS_2} to yield a satisfying approximation to $p_{\bm{y}}(\bm{\theta})$ independent of the parametric model choice.
This problem can be overcome by tempering the likelihood as described in the following section.
\subsection{Tempering the likelihood}
In order to bridge the distance between prior and likelihood one may break down the single CE problem into several smaller ones. To this end, we define a sequence of PDFs $\{q_t(\bm{\theta})\}_{j=1}^m$ with
\begin{equation}
q_t(\bm{\theta}) \vcentcolon  =  \frac{L^{\beta_t}(\bm{\theta})  p_0(\bm{\theta})}{Z_t}, 
\end{equation}
where $Z_t = \int_{\mathcal{X}} L^{\beta_t}(\bm{\theta})  p_0(\bm{\theta}) \mathrm{d}\bm{\theta}$ and ensuring $0 = \vcentcolon   \beta_0 < \beta_1 < \dots < \beta_{m-1} < \beta_{m} \vcentcolon  = 1$ such that $q_0(\bm{\theta}) \vcentcolon  = p_0(\bm{\theta})$ and $q_m(\bm{\theta}) \vcentcolon  = p_{\bm{y}}(\bm{\theta})$. The idea is to start with samples from $p_0(\bm{\theta})$ and select $\beta_1$ small enough to facilitate an accurate estimate $\widehat{\bm{v}}_1$. Next, upon selecting $\beta_2 \in (\beta_1, 1]$, samples from $q(\bm{\theta},\widehat{\bm{v}}_1)$ can be used to estimate $\widehat{\bm{v}}_2$. This procedure is repeated until $\beta_m = 1$ after $m$ steps and the CE problem is solved for the target posterior density. The $t$-th CE minimization problem reads
\begin{equation}
\label{eq:CE_IS_3}
\widehat{\bm{v}}_t = \argmax\limits_{\bm{v} \in \mathcal{V}} \frac{1}{n} \sum\limits_{k=1}^{n}   \ln(q(\bm{\theta}_k,\bm{v})) w_t(\bm{\theta}_k),~~~\bm{\theta}_k \stackrel{i.i.d.}{\sim} q(\bm{\theta},\widehat{\bm{v}}_{t-1}),
\end{equation}
with $w_t(\bm{\theta}) = L^{\beta_t}(\bm{\theta}) p_0(\bm{\theta})/q(\bm{\theta},\widehat{\bm{v}}_{t-1})$. 
\\~\\
In \cref{eq:CE_IS_3}, the likelihood ratio or weight $w_t(\bm{\theta})$ accounts for the fact that the $t$-th PDF parameter estimate $\widehat{\bm{v}}_{t}$ is based on samples from the $(t-1)^{\mathrm{th}}$ PDF $q(\bm{\theta},\widehat{\bm{v}}_{t-1})$. The variance of $\widehat{\bm{v}}_t$ depends on the variance of the weights $w_t(\bm{\theta})$. In particular, if the numerator PDF of $w$ has fatter tails than its denominator PDF, the weight variance blows up and the parameter estimate $\widehat{\bm{v}}_t$ deteriorates.
The normalized \emph{effective sample size} (nESS) is a common performance metric of IS that is directly related to the variance of the weights \cite[Chapter 9]{Owen2013}:
\begin{equation}
n_{\mathrm{eff}} = \frac{1}{1 + \delta_w^2}~~~\mathrm{with}~~~\delta_w = \frac{\sqrt{\mathbb{V}[w(\bm{\Theta})]}}{\mathbb{E}[w(\bm{\Theta})]}
\label{eq:nESS}
\end{equation}
the coefficient of variation of the weights.
Therefore in \cite{Engel2021}, $\beta_t$ is computed adaptively in each step such as to achieve a target nESS $n_{\mathrm{eff}}^*$ by utilizing a sample-based estimate of the coefficient of variation of the weights $\widehat{\delta}_w$:
\begin{equation}
\label{eq:compute_betat}
\beta_t = \argmin\limits_{\beta_{j-1} < \beta \le 1} \left(n_{\mathrm{eff}}^* - \frac{1}{1 + \widehat{\delta}_w(\beta)^2}\right)^2 = \argmin\limits_{\beta_{j-1} < \beta \le 1} \left(n_{\mathrm{eff}}^* - \frac{(\sum_{k=1}^n w(\bm{\theta}_k;\beta))^2}{\sum_{k=1}^n w^2(\bm{\theta}_k;\beta)}\right)^2.
\end{equation}
The weights on the right-hand side of \cref{eq:compute_betat} can be evaluated approximately by assuming $q(\bm{\theta},\widehat{\bm{v}}_{t-1}) = q_{t-1}(\bm{\theta})$ in each step, such that $w(\bm{\theta}) \propto L(\bm{\theta})^{\beta_t - \beta_{t-1}}$ (the factor $Z_{t-1}$ required here for equality cancels out in \cref{eq:compute_betat} and is immaterial to its solution).
\subsection{Method}
 In \cite{Engel2021}, CEBU is implemented in the $d$-dimensional standard-normal space $(\mathcal{U},\mathcal{B}(\mathcal{U}),\mathbb{P}_{\bm{U}})$ with $\mathcal{U} = \mathbb{R}^d$, so that the standard-normal random vector $\bm{U} \sim \varphi_d(\bm{u})$, where $\varphi_d$ denotes the $d$-dimensional standard-normal PDF.
Under a suitable isoprobabilistic transformation $T: \bm{\Theta} \rightarrow \bm{U}$, e.g., using the inverse CDF transform, the Rosenblatt transform \cite{Rosenblatt1952} or copula models \cite{Liu1986,Nelsen2010,Torre2019}, arbitrary priors $p_0(\bm{\theta})$ are transformed to $\varphi_d(\bm{u})$ while the likelihood in standard-normal space is given as $\tilde{L} \vcentcolon = (L \circ T^{-1})(\bm{u})$. Similarly, we define 
$\tilde{p}_{\bm{y}} \vcentcolon =\tilde{L}(u) \varphi(\bm{u})/Z$ where the evidence $Z$ is invariant under the transformation $T$ \cite[Appendix A]{Engel2021}.
\\~\\
The CEBU loop terminates once $\beta_t = 1$. Then, a set of $n$ samples is drawn from the final parametric density corresponding to $\beta_t = 1$. These samples are subsequently reweighted to generate samples from the true posterior distribution $\tilde{p}_{\bm{y}}$. To this end, a final set of weights $\bm{\mathrm{w}}_{\mathrm{final}}(\bm{u}) = \tilde{L}(\bm{u})\varphi_d(\bm{u})/q(\bm{u},\widehat{\bm{v}}_{t})$ is computed as the likelihood ratio of the unnormalized posterior in standard-normal space $\tilde{L}(\bm{u})\varphi_d(\bm{u})$ and the parametric density corresponding to $\beta_t = 1$, $q(\bm{u},\widehat{\bm{v}}_{t})$. The evidence can be written as 
\begin{equation}
Z = \mathbb{E}_{\varphi_d}[\tilde{L}(\bm{U})] = \mathbb{E}_{q(\bm{u},\widehat{\bm{v}}_{t})}[\tilde{L}(\bm{U})\varphi_d(\bm{U})/q(\bm{U},\widehat{\bm{v}}_{t})] = \mathbb{E}_{q(\bm{u},\widehat{\bm{v}}_{t})}[\bm{\mathrm{w}}_{\mathrm{final}}],
\end{equation}
which suggests estimating $Z$ as
\begin{equation}
\widehat{Z} = \frac{1}{n} \sum_{k=1}^n w_{\mathrm{final}}(\bm{u}_k),~~~\bm{u}_k \stackrel{i.i.d.}{\sim} q(\bm{u},\widehat{\bm{v}}_{t}).
\end{equation}
A desired number of $N$ weighted posterior samples $\{\bm{u}_k\}_{k=1}^N$ may then be obtained by resampling the last set of $n$ samples corresponding to $\beta_t = 1$ with replacement and weighted with the normalized final weights $\{\bm{\mathrm{w}}_{\mathrm{final}} (\bm{u}_k)/(n\widehat{Z})\}_{k=1}^n$. In \cite{Engel2021}, the authors use a stratified version of this resampling step based on \cite{Elvira2017}.
In a final step, these samples are transformed back to $\bm{\Theta}$-space through applying the inverse transform $T^{-1}$.
The entire procedure is summarized in \cref{alg:cebu}.
\begin{algorithm}[H]
	\caption{CE-BU}\label{alg:cebu}
	\hspace*{\algorithmicindent}\textbf{Input} Likelihood $L$, transform $T$, target nESS $n_{\mathrm{eff}}^*$,  \# post. samples $N$, \# samples/level $n$\\
	\hspace*{\algorithmicindent}\textbf{Output} posterior samples $\mathbf{\Theta}_{\mathrm{post}}$, estimated evidence $\widehat{Z}$
	\begin{algorithmic}[1]
		\STATE Set $t \gets 0$, $\beta_0 \gets 0$, $\widehat{\bm{v}}_0$ (so that $q_0(\bm{u}) = \varphi_d(\bm{u})$)
		\WHILE{$\beta_t < 1$}
		\STATE $t \gets t+1$
		\STATE Sample $\bm{\mathrm{U}} \in \mathbb{R}^{n \times d} \gets \{\bm{u}_k \stackrel{i.i.d.}{\sim} q(\bm{u},\widehat{\bm{v}}_{t-1})\}_{k = 1}^n $ 
		\STATE Evaluate $\bm{\mathrm{\ell}}  \in \mathbb{R}^{n\times 1} \gets (L \circ T^{-1})(\bm{\mathrm{U}})$
		\STATE Compute $\beta_{t}$ with \cref{eq:compute_betat}
		\STATE Evaluate $\bm{\mathrm{w}} \in \mathbb{R}^{n\times 1} \gets \bm{\mathrm{\ell}}^{\beta_t - \beta_{t-1}} $
		\STATE Compute $\widehat{\bm{v}}_{t}$ with \cref{eq:CE_IS_3}
		\ENDWHILE
		\STATE Sample $\bm{\mathrm{U}} \in \mathbb{R}^{n \times d} \gets \{\bm{u}_k \stackrel{i.i.d.}{\sim} q(\bm{u},\widehat{\bm{v}}_{t})\}_{k = 1}^n $ 
		\STATE Evaluate $\bm{\mathrm{w}}_{\mathrm{final}} \in \mathbb{R}^{n\times 1} \gets \left\{\frac{ (L\circ T^{-1})(\bm{u}_k)\varphi_d(\bm{u}_k)}{q(\bm{u}_k,\widehat{\bm{v}}_{t})}\right\}_{k = 1}^n $
		\STATE Estimate evidence $\widehat{Z} \gets \frac{1}{n} \sum_{k=1}^n \mathrm{w}_{\mathrm{final},k}$
		\STATE Normalize weights $\bar{\bm{\mathrm{w}}}_{\mathrm{final}} \gets \bm{\mathrm{w}}_{\mathrm{final}}/(n \widehat{Z})$ 
		\STATE $\mathbf{U}_{\mathrm{post}} \gets$ Resample (with replacement) $N$ times from $\mathbf{U}$ with weighting $\bar{\bm{\mathrm{w}}}_{\mathrm{final}}$
		\STATE $\mathbf{\Theta}_{\mathrm{post}} \gets T^{-1}(\mathbf{U}_{\mathrm{post}})$ 
		\RETURN{$\mathbf{\Theta}_{\mathrm{post}}$, $\widehat{Z}$}
	\end{algorithmic}
\end{algorithm}
\section{CEBUred: Certified dimension reduction for CEBU}
\label{sec:cebured}
\subsection{Linear subspaces of \texorpdfstring{$\mathcal{U}$}{U}}
\cite{Engel2021} test both Gaussian and vMFN mixture models for $q$. A $K$-component Gaussian mixture requires fitting $Kd(d+3)/2+K-1$ parameters whereas a $K$-component vMFN mixture features only $K(d+3)+K-1$ parameters. In spite of the more advantageous linear scaling in $d$ offered by vMFN mixtures, the required number of samples per CE-level can quickly exceed the computational budget if $d$ is large. 

In \cite{Engel2021}, it is shown that the CE method with the Gaussian mixture model is able to obtain accurate representations of posterior densities in various problem settings. However, the Gaussian mixture model performs poorly in high-dimensional IS \cite{Geyer2019}. This is due to the fact that IS weights with respect to Gaussian densities tend to degenerate in high dimensions. Further, the number of parameters of the GM model increases quadratically with the input dimension.
The latter implies that the required number of samples per CE-level to obtain accurate parameter estimates becomes prohibitively large in high dimensions.
To alleviate these problems, we draw on the ideas presented in \cite{Zahm2021} to determine a low-dimensional linear subspace of $\mathcal{X}$ in which an effective IS density can be constructed. The resulting approach can be viewed as an extension of the CE method with failure-informed dimension for rare event estimation, proposed in \cite{Uribe2020a}.
\\~\\
In each step of CEBU, for the tempered posterior distribution $\tilde{p}_{\bm{y},\beta}(\bm{u}) = {Z}^{-1}_\beta \tilde{L}^{\beta}(\bm{u}) \varphi_d(\bm{u})$ with \smash{${Z}_\beta = \int_{\mathbb{R}^d} \tilde{L}^{\beta}(\bm{u}) \varphi_d(\bm{u}) \mathrm{d}\bm{u}$} we seek an approximation of the form
\begin{equation}
\tilde{p}_{\bm{y},\beta}^{(r)} \propto (g \circ \proj)(\bm{u}) \varphi_d(\bm{u}),
\end{equation}
where $g: \mathbb{R}^d \rightarrow \mathbb{R}_{>0}$ is a Borel-measurable function referred to as \emph{profile function} in the following. $\proj \in \mathbb{R}^{d \times d}$ is a rank-$r$ projection matrix, i.e., $\proj \circ \proj = \proj$. Any $\bm{u} \in \mathbb{R}^d$ can be decomposed as 
$\bm{u} = \proj \bm{u} +  \projperp \bm{u} = \bm{u}_r + \bm{u}_{\perp}$  with the complementary projection $\projperp \vcentcolon = \mathbf{I}_d - \proj$ satisfying $\mathrm{Im}(\projperp) = \mathrm{Ker}(\proj)$. We call $\mathcal{U}_r \vcentcolon = \mathrm{Im}(\proj)$ the \emph{likelihood-informed subspace} (LIS) and $\mathcal{U}_\perp \vcentcolon = \mathrm{Im}(\projperp)$ the \emph{complementary subspace} (CS). The LIS and CS are at this point still subsets of the ambient space $\mathbb{R}^d$ so that no effective dimension reduction is achieved by their introduction. However, they correspond to lower-dimensional spaces we refer to as local LIS $\bar{\mathcal{U}}_r$ and local CS $\bar{\mathcal{U}}_\perp$, where in standard-normal space $\bar{\mathcal{U}}_r = \mathbb{R}^r$ and $\bar{\mathcal{U}}_\perp = \mathbb{R}^{d-r}$. We discuss the mapping to these local subspaces in more detail in \cref{sec:method}. The profile function $g$ is only a function of $\bm{u}_r \in \mathcal{U}_r$ and is constant in $\bm{u}_\perp \in \mathcal{U}_\perp$.  Following \cite{Uribe2020a}, we first define an optimal $g$ for the tempered posterior distributions of CEBU given a projection $\proj$ in \cref{sec:g}. Next, we identify the projection that minimizes the KLD between full and low-rank posterior in \cref{sec:Pr} and lay out the certified dimensionality reduction for CEBU in \cref{sec:method}.
\subsection{Optimal profile function \texorpdfstring{$g$}{g}}
\label{sec:g}
\cite{Zahm2021} show that for a given projection matrix $\proj$, the optimal profile function $g_\beta^\star(\bm{u})$ that minimizes the KLD $D_{\mathrm{KL}}(\tilde{p}_{\bm{y},\beta}||\tilde{p}_{\bm{y},\beta}^{(r)})$, is the following conditional expectation  
\begin{equation}
\label{eq:conditional_expectation}
\mathbb{E}_p[\tilde{L}^\beta(\bm{U}) |\proj \bm{u}]: \bm{u} \rightarrow \int_{\mathbb{R}^{d-r}} \tilde{L}^\beta(\proj \bm{u} + \bm{\Phi}_{\perp} \bar{\bm{u}}_{\perp})  p_{\perp}(\bar{\bm{u}}_{\perp}|\proj \bm{u}) \mathrm{d} \bar{\bm{u}}_{\perp},
\end{equation}
where $\bm{\Phi}_\perp \in \mathbb{R}^{d \times d-r}$ such that $\mathrm{span}(\bm{\Phi}_\perp) = \mathrm{Ker}(\proj)$ and $\bar{\bm{u}}_\perp \in \mathbb{R}^{d-r}$. The conditional PDF $p_{\perp}(\bar{\bm{u}}_{\perp}|\proj \bm{u})$ reads
\begin{equation}
\label{eq:conditional_density}
p_{\perp}(\bar{\bm{u}}_{\perp}|\proj \bm{u}) = \frac{\varphi_d(\proj \bm{u} + \bm{\Phi}_{\perp} \bar{\bm{u}}_{\perp})}{\int_{\mathbb{R}^{d-r}} \varphi_d(\proj \bm{u} + \bm{\Phi}_{\perp} \bar{\bm{u}}_{\perp}') \mathrm{d}\bm{u}_\perp'},
\end{equation}
which, by convention, equals zero whenever the denominator of \cref{eq:conditional_density} equals zero.
Following from the optimality of \cref{eq:conditional_expectation}, the optimal reduced posterior reconstruction in standard-normal space reads
\begin{equation}
\label{eq:optimal_posterior}
\tilde{p}_{\bm{y},\beta}^{(r,\star)} \propto \mathbb{E}_p[\tilde{L}^\beta(\bm{U}) |\proj \bm{u}] \varphi_d(\bm{u}).
\end{equation}
\cite{Zahm2021} also remarks that the conditional expectation \cref{eq:conditional_expectation} is not only optimal with respect to the KL divergence but also minimizes the mean-square reconstruction error of the likelihood function with respect to the prior measure $\mathbb{E}_p[(\tilde{L}^\beta(\bm{U}) - (g \circ \proj)(\bm{U}))^2]$.
\subsection{Optimal projection \texorpdfstring{$\proj$}{P}}
\label{sec:Pr}
Under assumptions on the prior distribution that hold in the standard-normal setting \cite[Example 2.6]{Zahm2021}, the \emph{subspace logarithmic Sobolev inequality} in \cite[Theorem 2.9]{Zahm2021} states that $\int_{\mathbb{R}^d} \lVert  \nabla h(\bm{u}) \lVert^2 \varphi_d(\bm{u}) \mathrm{d}\bm{u} \leq \infty$ for any continuously differentiable function $h: \mathbb{R}^d \rightarrow  \mathbb{R^d}$ and for any projection $\proj \in \mathbb{R}^{d \times d}$,
\begin{equation}
\label{eq:subspace_inequality}
\int_{\mathbb{R}^d} h^2(\bm{u}) \ln \left( \frac{h^2(\bm{u})}{\mathbb{E}_{\varphi_d}[h(\bm{U}) | \proj \bm{u}]} \right) \varphi_d(\bm{u}) \mathrm{d}\bm{u} \leq 2 \int_{\mathbb{R}^d} \lVert (\mathbf{I} - \proj^\tran) \nabla h(\bm{u}) \lVert^2 \varphi_d(\bm{u}) \mathrm{d}\bm{u}.
\end{equation}
By choosing \smash{$h^2(\bm{u}) = Z_\beta^{-1} \tilde{L}^\beta (\bm{u})$} we obtain the KLD \smash{$D_{\mathrm{KL}}(\tilde{p}_{\bm{y},\beta} || \tilde{p}_{\bm{y},\beta}^{(r,\star)})$} on the left-hand side of \cref{eq:subspace_inequality}. With \smash{$\nabla h(\bm{u}) = \frac{1}{2}(Z_\beta^{-1} \tilde{L}^\beta (\bm{u}))^{\frac{1}{2}} \nabla \ln \tilde{L}^\beta (\bm{u})$}, an upper bound on \smash{$D_{\mathrm{KL}}(\tilde{p}_{\bm{y},\beta} || \tilde{p}_{\bm{y},\beta}^{(r,\star)})$} emerges on the right-hand side of \cref{eq:subspace_inequality} as
\begin{align*}
\label{eq:upper_bound_kl}
D_{\mathrm{KL}}(\tilde{p}_{\bm{y},\beta} || \tilde{p}_{\bm{y},\beta}^{(r,\star)}) & \leq  \frac{1}{2} \int_{\mathbb{R}^d} \lVert (\mathbf{I} - \proj^\tran) \nabla \ln \tilde{L}^\beta (\bm{u}) \lVert^2 \tilde{p}_{\bm{y},\beta}(\bm{u}) \mathrm{d}\bm{u}\\
&=\frac{1}{2} \int_{\mathbb{R}^d} \mathrm{tr}\left[(\mathbf{I} - \proj^\tran) \beta^2 \nabla \ln \tilde{L} (\bm{u}) (\nabla \ln \tilde{L} (\bm{u}))^\tran (\mathbf{I} - \proj)\right]  \tilde{p}_{\bm{y},\beta}(\bm{u}) \mathrm{d}\bm{u}\\
&=\frac{1}{2}\mathrm{tr}\left[(\mathbf{I} - \proj^\tran) \mathbf{H} (\mathbf{I} - \proj)\right]  =\vcentcolon \frac{1}{2} \mathcal{R}(\proj, \mathbf{H}),
\end{align*}
where $\mathcal{R}(\proj, \mathbf{H})$ is the mean-squared error incurred by approximating $\nabla \ln \tilde{L}(\bm{U})$ with $\proj^\tran \nabla \ln \tilde{L}(\bm{U})$ when $\bm{U} \sim \tilde{p}_{\bm{y},\beta}(\bm{u})$ and we define
\begin{equation}
\mathbf{H} \vcentcolon= \beta^2 \mathbb{E}_{\tilde{p}_{\bm{y},\beta}}\left[\nabla \ln \tilde{L} (\bm{U}) (\nabla \ln \tilde{L} (\bm{U}))^\tran \right].
\end{equation}
Our goal is to find the rank-r projection that minimizes $\mathcal{R}(\proj, \mathbf{H})$.
\cite[Proposition 2.11]{Zahm2021} states that a minimizer of $\mathcal{R}(\proj, \mathbf{H})$ over all viable projections of rank $r$ is given by the $r$ eigenvectors of $\mathbf{H}$ corresponding to its $r$ leading eigenvalues. Let the solutions of the eigenproblem $\mathbf{H}\bm{\phi}_i =  \bm{\phi}_i \lambda_i$, $\{\bm{\phi}_i,\lambda_i\}_{i=1}^d$, be ordered so that $\lambda_1 \geq \lambda_2 \geq \cdots \geq \lambda_{d}$, then collecting $\bm{\Phi}_r \vcentcolon = [\bm{\phi}_1, \bm{\phi}_2, \dots, \bm{\phi}_r] \in \mathbb{R}^{d \times r}$, the optimal projector is given as $\proj = \bm{\Phi}_r \bm{\Phi}_r^\tran$. With this definition of the projection and since the standard-normal prior satisfies inequality \cref{eq:subspace_inequality}, the accuracy of the reduced posterior can be controlled using a tolerance $\epsilon$ as
\begin{equation}
\label{eq:criterion}
D_{\mathrm{KL}}(\tilde{p}_{\bm{y},\beta} || \tilde{p}_{\bm{y},\beta}^{(r,\star)}) \leq \frac{\beta^2}{2}\sum_{i=r+1}^d \lambda_i \leq \epsilon.
\end{equation}
Upon selecting $\epsilon$ and computing the eigenpairs of $\mathbf{H}$, we choose $r$ as small as possible so that \cref{eq:criterion} holds. Efficient dimension reduction is therefore contingent on a sharp decay of the $\mathbf{H}$-spectrum, which is a property of the computational model $f$ and the observation model $p(\bm{u},\bm{y})$ (i.e., prior and likelihood). 
\subsection{Method}
\label{sec:method}
\begin{figure}[!ht]
	\centering
	\def\svgwidth{1\textwidth}
	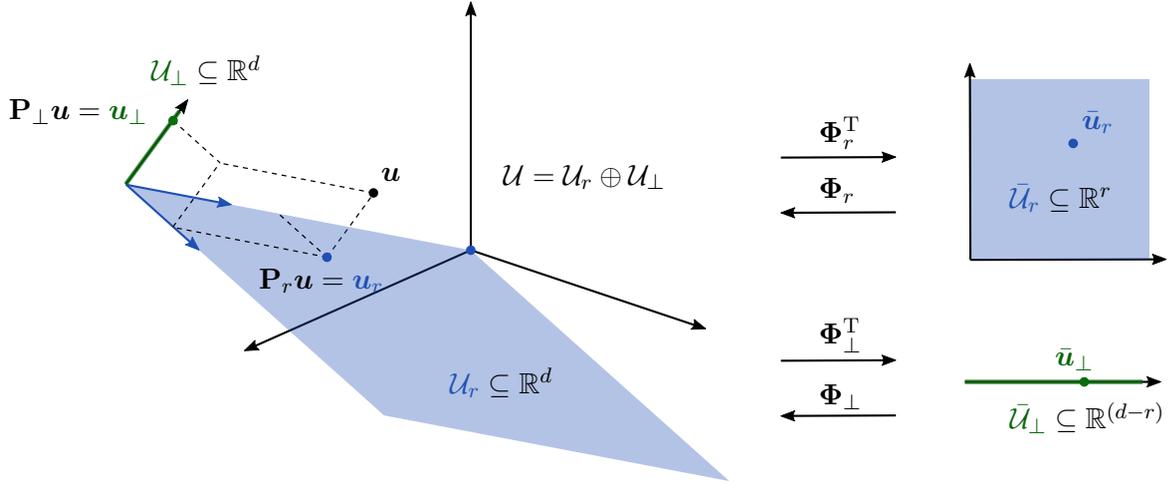{}
	\caption{Ambient space $\mathbb{R}^d$ along with the LIS $\mathcal{U}_r$ and CS $\mathcal{U}_\perp$ as defined by $\proj$ and $\projperp$ (left) and their local counterparts $\bar{\mathcal{U}}_r$ and $\bar{\mathcal{U}}_\perp$ (right).}
	\label{fig:map}
\end{figure}

\begin{figure}[!ht]
	\centering
	\def\svgwidth{1\textwidth}
	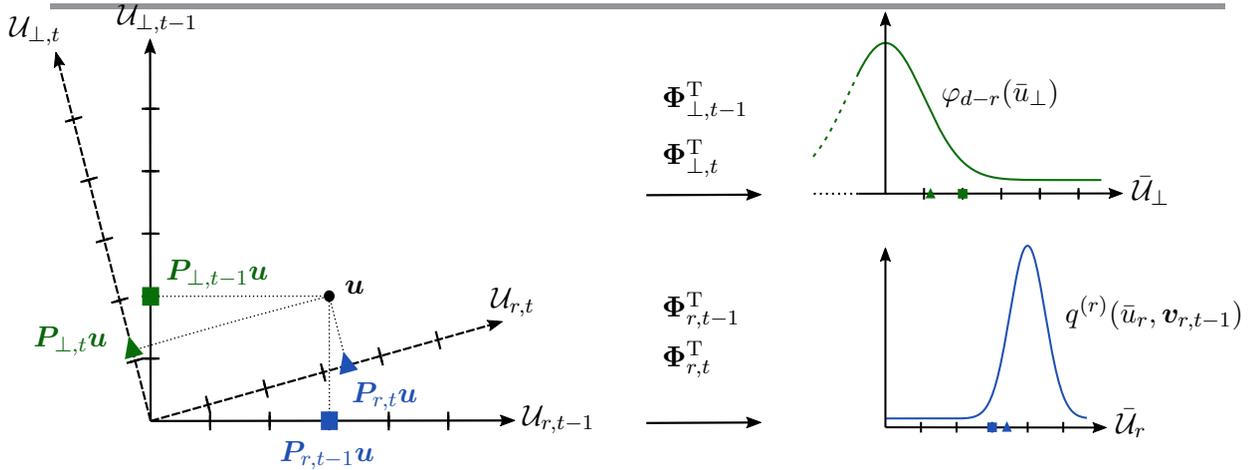{}
	\caption{Left: Mapping between two subsequent orthogonal bases $\bm{\Phi}_{t-1}$ and $\bm{\Phi}_t$ in ambient space. Right: Mapping from the two ambient bases to the local LIS (upper right) and CS (lower right).}
	\label{fig:paramTrans}
\end{figure}

$\bm{\Phi}_r$ is the eigenspace of the symmetric matrix $\mathbf{H}$ and thus is an orthogonal basis of $\mathcal{U}_r$. $\bm{\Phi}_r$ maps the ambient LIS coordinate $\bm{u}_r \in \mathcal{U}_r$ to its local counterpart $\bar{\bm{u}}_r \in \bar{\mathcal{U}}_r = \mathbb{R}^{r}$ as
$\bar{\bm{u}}_r = \bm{\Phi}_r^\tran \bm{u}_r$. In the same way, we define $\bm{\Phi}_\perp \vcentcolon = [\bm{\phi}_{r+1}, \bm{\phi}_{r+2}, \dots, \bm{\phi}_d] \in \mathbb{R}^{d \times d-r}$, so that the ambient CS coordinate $\bm{u}_\perp$ is mapped to its local counterpart $\bar{\bm{u}}_\perp \in \bar{\mathcal{U}}_\perp = \mathbb{R}^{d-r}$ as $\bar{\bm{u}}_\perp = \bm{\Phi}_\perp^\tran \bm{u}_\perp$.
Thus, we can write any $\bm{u} \in \mathbb{R}^d$ as 
$\bm{u} = \bm{\Phi}_r \bar{\bm{u}}_r + \bm{\Phi}_\perp \bar{\bm{u}}_\perp$
and collect the ambient coordinate of $\mathbb{R}^d$ with respect to the basis defined by $\bm{\Phi} = [\bm{\Phi}_r,\bm{\Phi}_\perp]$ as
\begin{equation}
\bar{\bm{u}}=\left[\begin{matrix}\bar{\bm{u}}_r\\\bar{\bm{u}}_\perp\end{matrix}\right] = \bm{\Phi}^\tran \cdot \bm{u} =
\left[\begin{matrix}\bm{\Phi}_r^\tran\\ \bm{\Phi}_\perp^\tran\end{matrix}\right] \cdot \bm{u}. 
\end{equation}
The LIS and CS in local and ambient coordinates are illustrated in \cref{fig:map}. 
Due to orthogonality of $\bm{u}_r$ and $\bm{u}_\perp$ and the rotatinal symmetry of the standard-normal PDF, we may factorize the prior as $\varphi_d(\bm{u}) = \varphi_r(\bar{\bm{u}}_r) \varphi_{d-r}(\bar{\bm{u}}_\perp)$. With this local coordinate prior, the reduced tempered posterior \cref{eq:optimal_posterior} reads
\begin{equation}
\label{eq:factorized_posterior}
p_{\bm{y},\beta}^{(r,\star)}(\bar{\bm{u}}) \propto \underbrace{\mathbb{E}_{\varphi_d}[\tilde{L}^{\beta_t}(\bm{U}) |\bm{\Phi}_r \bar{\bm{u}}_r] \varphi_r(\bar{\bm{u}}_r)}_{\substack{\mathrm{reduced~tempered}\\\mathrm{posterior}}} \underbrace{\varphi_{d-r}(\bar{\bm{u}}_\perp).}_{\substack{\mathrm{complementary}\\\mathrm{prior}}}
\end{equation}
By analogy with CEBU, in the $t$-th step of CEBUred, we approximate the reduced tempered posterior in \cref{eq:factorized_posterior} with a parametric model $q^{(r)}(\bar{\bm{u}}_r,\bm{v}_{r,t}): \bar{\mathcal{U}}_r \rightarrow \mathbb{R}_{>0}$. The parametric, tempered posterior is
\begin{equation}
\label{eq:factorized_parametric:posterior}
q(\bar{\bm{u}},\bm{v}_{r,t}) = q^{(r)}(\bar{\bm{u}}_r,\bm{v}_{r,t}) \varphi_{d-r}(\bar{\bm{u}}_\perp).
\end{equation}
Following \cite{Uribe2020a}, we select a Gaussian model for $q^{(r)}(\bar{\bm{u}}_r,\bm{v}_{r,t})$, although more complicated PDFs such as mixture models may be used as well here. 
The parameter set $\bm{v}_{r,t}=\{\bm{\mu}_{r,t} \in \mathbb{R}^r,\bm{\Sigma}_{r,t} \in \mathbb{R}^{r \times r}\}$ contains the mean vector $\bm{\mu}_{r,t}$ and covariance matrix $\bm{\Sigma}_{r,t}$ of the Gaussian model.
\\~\\
In the $t$-th step of CEBUred, the new temperature $\beta_t$ is computed according to \cref{eq:compute_betat}. If $t>1$, the likelihood in ambient space is evaluated by plugging samples from the previous' step's reduced biasing density $\bar{\bm{u}}_{r,k} \sim q^{(r)}(\bar{\bm{u}}_r,\bm{v}_{r,t})$ and the complementary prior $\bar{\bm{u}}_{\perp,k} \sim \varphi_{d-r}(\bar{\bm{u}}_\perp)$ in $\bm{u}_k = \bm{\Phi}_{r,t-1}\bar{\bm{u}}_{r,k} + \bm{\Phi}_{\perp,t-1} \bar{\bm{u}}_{\perp,k}$.
Thereafter, the gradient covariance matrix
$\mathbf{H}$ 
of the likelihood function with respect to the tempered posterior is estimated to determine
the current LIS and CS projections 
In each step but the first ($t>1$), a self-normalized IS estimate of $\mathbf{H}_t$ based on samples from the previous biasing density $q(\bar{\bm{u}},\widehat{\bm{v}}_{t-1})$ is computed as
\begin{equation}
\label{eq:H_estimate}
\widehat{\mathbf{H}}_t = \frac{\beta_t^2 \sum_{k=1}^{n_{\mathbf{H}}} \nabla \ln L(\bm{u}_k) (\nabla \ln L(\bm{u}_k))^\tran w_t(\bar{\bm{u}}_{k})}{\sum_{k=1}^{n_{\mathbf{H}}} w_t(\bar{\bm{u}}_k)}~~~~~\begin{cases}
\bar{\bm{u}}_{r,k} \stackrel{i.i.d.}{\sim} q^{(r)}(\bar{\bm{u}}_{r},\bm{v}_{r,t-1})\\    
\bar{\bm{u}}_{\perp,k} \stackrel{i.i.d.}{\sim} \varphi_{d-r}(\bar{\bm{u}}_{\perp})
\end{cases}.
\end{equation}
If $t=1$, the weights equal 1 and samples are drawn from the $d$-dimensional prior in ambient space, $\varphi_d(\bm{u})$ by setting $\bm{\Phi}_r = \mathbf{I}_d$ and $\bm{\Phi}_\perp = \bm{0}_d$. For any $t>1$ the weights are computed as
\begin{equation}
\label{eq:H_weights}
w_t(\bar{\bm{u}}) = \frac{\tilde{L}^{\beta_t}(\bm{\Phi}_r \bar{\bm{u}}_r + \bm{\Phi}_\perp \bar{\bm{u}}_\perp)  \varphi_r(\bar{\bm{u}}_r) \varphi_{d-r}(\bar{\bm{u}}_\perp)}{q(\bar{\bm{u}},\bm{v}_{r,t-1})} = \frac{\tilde{L}^{\beta_t}(\bm{\Phi}_r \bar{\bm{u}}_r + \bm{\Phi}_\perp \bar{\bm{u}}_\perp)  \varphi_r(\bar{\bm{u}}_r)}{q^{(r)}(\bar{\bm{u}}_r,\bm{v}_{r,t-1})}.
\end{equation}
Upon computing the spectrum of $\widehat{\mathbf{H}}_t$, the LIS-dimension $r$ is selected according to \cref{eq:criterion}. Once the projections $\bm{\Phi}_r$ and $\bm{\Phi}_\perp$ are defined, the parameters of $q^{(r)}(\bar{\bm{u}}_r,\bm{v}_{r,t})$ are computed by minimizing the KLD $D_{\mathrm{KL}}(\tilde{p}_{\bm{y},\beta}^{(r,\star)}(\bar{\bm{u}})||q(\bar{\bm{u}},\bm{v}_t))$. As in \cref{eq:DKL}, this is equivalent to maximizing the negative cross-entropy between the two distributions, i.e.,
\begin{equation}\label{eq:reduced_parameters}
\begin{split}
\bm{v}_{r,t}
&= \argmin\limits_{\bm{v} \in \mathcal{V}} D_{\mathrm{KL}}\left(\tilde{p}_{\bm{y},\beta}^{(r,\star)} \lVert q(\cdot,\bm{v})\right)
 =\argmax\limits_{\bm{v} \in \mathcal{V}} - H\left(\tilde{p}_{\bm{y},\beta}^{(r,\star)}, q(\cdot,\bm{v})\right)\\
 & = \argmax\limits_{\bm{v}_r \in \mathcal{V}_r} \int_{\bar{\mathcal{U}}_r} \int_{\bar{\mathcal{U}}_\perp} \mathbb{E}_{\varphi_d}[\tilde{L}^{\beta_t}(\bm{U}) |\bm{\Phi}_r \bar{\bm{u}}_r] \ln \left(q^{(r)}(\bar{\bm{u}}_r,\bm{v}_r)\right)  \varphi_r(\bar{\bm{u}}_r) \varphi_{d-r}(\bar{\bm{u}}_\perp) \mathrm{d} \bar{\bm{u}}_\perp \mathrm{d} \bar{\bm{u}}_r \\
  & = \argmax\limits_{\bm{v}_r \in \mathcal{V}_r} \int_{\bar{\mathcal{U}}_r} \mathbb{E}_{\varphi_d}[\tilde{L}^{\beta_t}(\bm{U}) |\bm{\Phi}_r \bar{\bm{u}}_r]  \ln \left(q^{(r)}(\bar{\bm{u}}_r,\bm{v}_r)\right)  \varphi_r(\bar{\bm{u}}_r)  \mathrm{d} \bar{\bm{u}}_r \\
 & = \argmax\limits_{\bm{v}_r \in \mathcal{V}_r} \mathbb{E}_{\varphi_d}\left[\tilde{L}^{\beta_t}(\bm{\Phi}_r \bar{\bm{U}}_r + \bm{\Phi}_\perp \bar{\bm{U}}_\perp) \ln \left(q^{(r)}(\bar{\bm{u}}_r,\bm{v}_r)\right)\right]. 
\end{split}
\end{equation}
Throughout \cref{eq:reduced_parameters} the normalization constant $\tilde{Z}_t$ has been dropped as it is irrelevant for solving the optimization problem. The final equality in \cref{eq:reduced_parameters} is a consequence of the factorized prior in standard-normal space, i.e.,
\begin{equation}
\begin{split}
\mathbb{E}_{\varphi_d}[\tilde{L}^{\beta_t}(\bm{U}) |\bm{\Phi}_r \bar{\bm{u}}_r] 
&= \int_{\bar{\mathcal{U}}_\perp} \tilde{L}^{\beta_t}(\bm{\Phi}_r \bar{\bm{u}}_r + \bm{\Phi}_\perp \bar{\bm{u}}_\perp) \frac{\varphi_r(\bar{\bm{u}}_r) \varphi_{d-r}(\bar{\bm{u}}_{\perp})}{\int_{\mathbb{R}^{d-r}} \varphi_r(\bar{\bm{u}}_r) \varphi_{d-r}(\bar{\bm{u}}_\perp') \mathrm{d}\bar{\bm{u}}_\perp'} \mathrm{d}\bar{\bm{u}}_\perp\\
&= \int_{\bar{\mathcal{U}}_\perp} \tilde{L}^{\beta_t}(\bm{\Phi}_r \bar{\bm{u}}_r + \bm{\Phi}_\perp \bar{\bm{u}}_\perp) \varphi_{d-r}(\bar{\bm{u}}_{\perp}) \mathrm{d}\bar{\bm{u}}_\perp.
\end{split}
\end{equation}
An IS estimate of $\bm{v}_{r,t}$ based on samples from $q(\bar{\bm{u}},\widehat{\bm{v}}_{r,t-1})$ reads
\begin{equation}
\label{eq:CEBUred_parameter_estimate}
\widehat{\bm{v}}_{r,t} = \argmax\limits_{\bm{v}_r \in \mathcal{V}_r} \frac{1}{n} \sum\limits_{k=1}^{n}   \ln \left(q^{(r)}(\bar{\bm{u}}_{r,k},\bm{v}_r)\right) w_{t,\text{adj}}(\bar{\bm{u}}_{r,k},\bar{\bm{u}}_{\perp,k}),~~~\begin{cases}
\bar{\bm{u}}_{r,k} \stackrel{i.i.d.}{\sim} q^{(r)}(\bar{\bm{u}}_{r},\bm{v}_{r,t-1})\\    
\bar{\bm{u}}_{\perp,k} \stackrel{i.i.d.}{\sim} \varphi_{d-r}(\bar{\bm{u}}_{\perp})\end{cases}
\end{equation}
and requires the computation of the \textit{adjusted weights} $w_{t,\text{adj}}$
\begin{equation}
\label{eq:adjusted_weights}
w_{t,\text{adj}}(\bar{\bm{u}}_{r},\bar{\bm{u}}_{\perp}) = \frac{\tilde{L}^{\beta_t}(\bm{\Phi}_{r,t} \bar{\bm{u}}_r + \bm{\Phi}_{\perp,t} \bar{\bm{u}}_\perp)  \varphi_r(\bar{\bm{u}}_r) \varphi_{d-r}(\bar{\bm{u}}_\perp)}{q^{(d)}(\bar{\bm{u}},\widehat{\bm{v}}_{t,\text{adj}})}.
\end{equation}
Therein, $\widehat{\bm{v}}_{t,\text{adj}} = \{\bm{\mu}_{t,\text{adj}} \in \mathbb{R}^d ,\bm{\Sigma}_{t,\text{adj}} \in \mathbb{R}^{d \times d} \}$ represents the parameters of the $d$-dimensional Gaussian density $q(\bar{\bm{u}},\widehat{\bm{v}}_{t-1})$ expressed with respect to the updated orthogonal basis $\bm{\Phi}_t$. Computing adjusted weights with the transformed parameters is necessary to address non-matching bases in the numerator and denominator of \cref{eq:adjusted_weights}. That is, if the new basis $\bm{\Phi}_t$ differs from the basis $\bm{\Phi}_{t-1}$, the complementary prior will no longer be standard-normal with respect to $\bm{\Phi}_t$. The transformation from $\bm{\Phi}_{t-1}$ to $\bm{\Phi}_t$ is linear whereby $q(\bar{\bm{u}},\widehat{\bm{v}}_{t,\text{adj}})$ is Gaussian again and its parameters with respect to $\bm{\Phi}_t$ can be expressed as 
\begin{equation}
\label{eq:cotrafo}
	\bm{\mu}_{t,\text{adj}} = \bm{\Phi}_{t}^\tran \underbrace{\bm{\Phi}_{r,t-1} \bm{\mu}_{r,t-1}}_{\bm{\mu}_{t-1}}	,~~~ 
	\bm{\Sigma}_{t,\text{adj}} = \bm{\Phi}_{t}^\tran \underbrace{[\bm{\Phi}_{r,t-1} \bm{\Sigma}_{r,t-1} \bm{\Phi}_{r,t-1}^\tran + \bm{\Phi}_{\perp,t-1} \bm{\Phi}_{\perp,t-1}^\tran]}_{\bm{\Sigma}_{t-1}} \bm{\Phi}_{t}.
\end{equation}
$\bm{\mu}_{t-1}$ and $\bm{\Sigma}_{t-1}$ are the mean and covariance vector in ambient space that are subsequently transformed to the reduced spaces given the novel basis $\bm{\Phi}_t$. This transformation between local and global and subsequent subspaces in steps $t-1$ and $t$ is illustrated in \cref{fig:paramTrans}.

\subsection{Choosing \texorpdfstring{$n_{\mathbf{H}}$}{nH} and \texorpdfstring{$n$}{n} adaptively}
\label{sec:adaptive_sample_size}
$n_{\mathbf{H}}$ is the number of log-likelihood gradient evaluations used to compute $\widehat{H}_t$ in \cref{eq:H_estimate}. $n$ on the other hand is the number of direct evaluations of the tempered likelihood used to estimate the parameters of the $t$-th biasing density in \cref{eq:CEBUred_parameter_estimate}. In the absence of $f$-solvers that are specifically geared towards efficient gradient evaluation such as adjoint solvers \cite{Arora1979}, computing $\nabla \ln L(\bm{u})$ is considerably more expensive than evaluating $ L(\bm{u})$. 
\\~\\
\cite{Constantine2015} suggests a heuristic for determining $n_{\mathbf{H}}$ when estimating the second-moment matrix of the gradient $\nabla f$ of a Lipschitz-continuous function $f$, i.e., $\nabla f \leq a$, in order to discover an \emph{active subspace} of $f$. They use work of \cite{Gittens2011} on the spectrum of sums of ($n_{\mathbf{H}}$) random matrices to establish bounds on the relative accuracy of the estimated spectrum of $\mathbf{C} = \mathbb{E}_p[\nabla f(\nabla f)^\tran]$. In the context of CEBUred, we have $f = \beta \log \tilde{L}$ and $p = \tilde{p}_{\bm{y},\beta}$.
\\~\\
\cite[Corollary 3.5]{Constantine2015} states that for $\varepsilon \in (0, 1]$, $\mathbb{P}[|\widehat{\lambda}_r - \lambda_r|/\lambda_r \leq \varepsilon] \leq 2d^{-b}$ if the spectrum of $\widehat{\mathbf{H}}$, $\{\widehat{\lambda}_i\}_{i=1}^d$, is computed with $n_{\mathbf{H}} \ge 4 (b+1) a \lambda_1 \ln(d) / (\lambda_r \varepsilon)^2$ log-likelihood gradient samples.
Drawing on a matrix Bernstein inequality in \cite{Tropp2012}, \cite[Corollary 3.8]{Constantine2015} states that for $\varepsilon \in (0, 1]$, $\mathbb{P}[\mathbf{H} - \widehat{\mathbf{H}} \lVert_2 / \lVert \mathbf{H} \lVert_2 \leq \varepsilon] \leq  2 m^{1-3c/8}$ (the $2$-norm of a matrix here is its spectral norm, which also corresponds to its largest singular value) when estimating $\widehat{\mathbf{H}}$ with at least $n_{\mathbf{H}} \ge c a^2 \ln(d) / (\lambda_1 \varepsilon^2)$ samples. 
Finally, choosing $\varepsilon$ such that $\varepsilon \leq (\lambda_r - \lambda_{r+1})/(5 \lambda_1)$ and using this last lower bound on $n_{\mathbf{H}}$, the distance between the image of the local estimated and true LIS projections is bounded with high probability as well: the distance as measured with the spectral norm $d(\mathrm{Im}(\bm{\Phi}_r),\mathrm{Im}(\widehat{\bm{\Phi}}_r)) = \lVert \bm{\Phi}_r\bm{\Phi}_r^\tran - \widehat{\bm{\Phi}}_r \widehat{\bm{\Phi}}_r^\tran\lVert_2 = \lVert \bm{\Phi}_r^\tran \widehat{\bm{\Phi}}_\perp \lVert_2 $ is bounded as $\mathbb{P}[\lVert \bm{\Phi}_r^\tran \widehat{\bm{\Phi}}_\perp \lVert_2 \leq 4 \lambda_1 \varepsilon/(\lambda_r - \lambda_{r+1})] \leq 2 m^{1-3c/8}$ according to \cite[Corollary 3.10]{Constantine2015}.
\cite{Constantine2015}  translates this bound into a heuristic on account of $c$, $a$ and the true spectrum $\{\lambda_i\}_{i=1}^d$ being unknown in many use cases involving numerical/simulation models $f$. The heuristic emerges by summarizing all unknown constants in a fudge factor $\alpha_{\mathbf{H}}$ resulting in
\begin{equation}
n_{\mathbf{H}} = \alpha_{\mathbf{H}} r \ln (d),
\label{eq:n_H}
\end{equation}
where \cite{Constantine2015} recommends $\alpha_{H} \in [2,10]$ and the target rank $r$ corresponds to the smallest eigenvalue $\lambda_r$ that shall be estimated with the desired relative accuracy $\varepsilon$. 
Remarkably, the effort scales logarithmically with the ambient space dimension $d$ suggesting that we can hope to estimate $\mathbf{H}$ with relatively few log-likelihood gradient samples even in very high-dimensional settings. As the target rank $r$ is not known a priori, we detail an iterative procedure to jointly determine $r$ and $n_{\mathbf{H}}$ in \cref{alg:adaptive_H}.
\\~\\
The required number of samples in each level of the CEBU procedure in turn depends on the adaptively selected LIS-rank $r$ through the number of parameters that have to be fitted in the Gaussian reduced biasing density $n_{\mathrm{par}}$. In particular, an $r$-variate Gaussian requires fitting $n_{\mathrm{par}} = r(r+3)/2$ parameters. To select the number of samples required to accurately estimate $\widehat{\bm{v}}_{r,t}$ in \cref{eq:CEBUred_parameter_estimate}, we use the following heuristic:
\begin{equation}
\label{eq:n_r}
n(r) =  {\underbrace{\vphantom{\frac{1}{2}}\alpha_{\mathrm{par}} }_{\substack{\mathrm{~fudge~}\\ \mathrm{factor}}}}~ {\underbrace{\frac{1}{2} r(r+3)}_{\substack{\mathrm{number~of}\\ \mathrm{parameters}}}}~
{\underbrace{\vphantom{\frac{1}{2}}(1 + \delta_w^2)}_{\substack{\mathrm{inverse}\\ \mathrm{nESS}}}},
\end{equation}
where \cite{Constantine2015} recommend to chose $\alpha_{\mathrm{par}} \in [2,10]$. In case an adjoint solver is used for $f$, the estimation of $\widehat{H}_t$ as in \cref{eq:H_estimate} will return $n_{\mathbf{H}}$ likelihood evaluations as a byproduct that can be utilizied in estimating $\widehat{\bm{v}}_{r,t}$  so that only an effective $n(r) - n_{\mathbf{H}}$ new samples need to be drawn and evaluated at each level.
\begin{algorithm}[!ht]
	\caption{adapt\_H}\label{alg:adaptive_H}
	\hspace*{\algorithmicindent}\textbf{Input} Likelihood and log-gradient $\tilde{L}$ and $\nabla \ln \tilde{L}$, reduced biasing density $q^{(r)}(\bar{\bm{u}}_r,\bm{v}_{r})$, local \hspace*{1.44cm} LIS \& CS projections $\bm{\Phi}_r$ \& $\bm{\Phi}_\perp$, fudge factor $\alpha_{\mathbf{H}}$, error tolerance $\epsilon$, temperature $\beta_0$\\
	\hspace*{\algorithmicindent}\textbf{Output} Subspace samples $\bar{\bm{\mathrm{U}}}_r, \bar{\bm{\mathrm{U}}}_\perp$,  Likelihood samples $\bm{\ell}$, local LIS \& CS projections $\bm{\Phi}_r$ \\\hspace*{1.85cm}  \& $\bm{\Phi}_\perp$, temperature $\beta$, LIS-rank $r$, \# of $\mathbf{H}$-samples $n_{\mathbf{H}}$
	\begin{algorithmic}[1]
		\STATE Set $r \gets 1$, $d \gets \mathrm{rowdim}(\bm{\Phi}_r)$, $n_{\mathbf{H}} \gets \alpha_{\mathbf{H}} \ln(d)$, $\Delta n \gets n_{\mathbf{H}}$, $\bar{\bm{\mathrm{U}}}_r,\bar{\bm{\mathrm{U}}}_\perp, \bm{\ell},d\bm{\ell} \gets [~]$
		\WHILE{$\Delta n > 0$}
		\STATE Sample $\bar{\bm{\mathrm{U}}}_{r,\mathrm{add}} \in \mathbb{R}^{\Delta n \times r} \gets \{\bar{\bm{u}}_{r,k} \stackrel{i.i.d.}{\sim} q^{(r)}(\bar{\bm{u}}_r,\bm{v}_{r})\}_{k = 1}^{\Delta n} $
		\STATE Sample \smash{$\bar{\bm{\mathrm{U}}}_{\perp,\mathrm{add}} \in \mathbb{R}^{\Delta n \times (d-r)} \gets \{\bar{\bm{u}}_{\perp,k} \stackrel{i.i.d.}{\sim} \varphi_{d-r}(\bar{\bm{u}}_\perp)\}_{k = 1}^{\Delta n}$}
		\STATE Append vertically $\bar{\bm{\mathrm{U}}}_r \gets [\bar{\bm{\mathrm{U}}}_r,\bar{\bm{\mathrm{U}}}_{r,\mathrm{add}}],~\bar{\bm{\mathrm{U}}}_\perp \gets [\bar{\bm{\mathrm{U}}}_\perp,\bar{\bm{\mathrm{U}}}_{\perp,\mathrm{add}}]$
		\STATE Compute $\bm{\ell}_{\mathrm{add}} \gets \tilde{L}(\bar{\bm{U}}_{r,\mathrm{add}}^\tran \bm{\Phi}_r  + \bar{\bm{U}}_{\perp,\mathrm{add}}^\tran \bm{\Phi}_\perp)$ and $ d\bm{\ell}_{\mathrm{add}} \gets \nabla \tilde{L}(\bar{\bm{U}}_{r,\mathrm{add}}^\tran \bm{\Phi}_r  + \bar{\bm{U}}_{\perp,\mathrm{add}}^\tran \bm{\Phi}_\perp)$
		\STATE Append vertically $\bm{\ell} \gets [\bm{\ell},\bm{\ell}_{\mathrm{add}}] ,d\bm{\ell} \gets [d\bm{\ell},d\bm{\ell}_{\mathrm{add}}]$
		\STATE Evaluate $\beta$ and $\mathbf{w}$ in function of $\beta_0$, $\bar{\bm{U}}_r$, $\bar{\bm{U}}_\perp$, $\bm{\ell}$, $q^{(r)}(\bar{\bm{U}}_r,\bm{v}_{r})$ with \cref{eq:compute_betat} \&  \cref{eq:H_weights}
		\STATE Evaluate \smash{$\widehat{\mathbf{H}}$} in function of $\beta$, $\mathbf{w}$ and $d \bm{\ell}$ with \cref{eq:H_estimate}
		\STATE Evaluate \smash{$\{\bm{\phi}_i,\lambda_i\}_{i=1}^d \gets \solve\{\bm{\phi} \in \mathbb{R}^d, \lambda \in \mathbb{R}: \widehat{\mathbf{H}}\bm{\phi} = \bm{\phi} \lambda\}$}
		\STATE Select $r$ in function of $\epsilon$ and $\{\bm{\phi}_i,\lambda_i\}_{i=1}^d$ with \cref{eq:criterion}

		\STATE Set $\Delta n = \alpha_{H} r \ln (d) - n_{\mathbf{H}}$
		\ENDWHILE
		\STATE Define $\bm{\Phi}_r \gets [\bm{\phi}_1,\dots,\bm{\phi}_r]$, $\bm{\Phi}_\perp \gets [\bm{\phi}_{r+1},\dots,\bm{\phi}_d]$
		\RETURN{$\bar{\bm{\mathrm{U}}}_r$, $\bar{\bm{\mathrm{U}}}_\perp$, $\bm{\ell}$, $\bm{\Phi}_r$, $\bm{\Phi}_\perp$, $\beta$, $r$, $n_{\mathbf{H}}$}
	\end{algorithmic}
\end{algorithm}
The CEBUred algorithm is summarized in \cref{alg:cebured}.
\begin{algorithm}[!ht]
	\caption{CE-BU-red}\label{alg:cebured}
	\hspace*{\algorithmicindent}\textbf{Input}  Likelihood and log-gradients $\tilde{L}$ and $\nabla \ln \tilde{L}$, transform $T$, parameters $n_{\mathrm{eff}}^*, \alpha_{\mathbf{H}}, \alpha_{\mathrm{par}} , \epsilon$,  \# post. samples $N$,\\
	\hspace*{\algorithmicindent}\textbf{Output} posterior samples $\mathbf{\Theta}_{\mathrm{post}}$, estimated evidence $\widehat{Z}$
	\begin{algorithmic}[1]
		\STATE Set $t \gets 0$, $\beta_0 \gets 0$, $\bm{\Phi}_{r,0} \gets \mathbf{I}_{d \times d}$, $\bm{\Phi}_{\perp,0} \gets 0$, $\widehat{\bm{v}}_{r,0} = \{\bm{0}_{d}, \mathbf{I}_{d \times d}\}$, 
		\WHILE{$\beta_t < 1$}
		\STATE $t \gets t+1$
		\STATE $\bar{\bm{\mathrm{U}}}_{r,0}, \bar{\bm{\mathrm{U}}}_{\perp,0}, \bm{\ell}_0,\bm{\Phi}_{r,t}, \bm{\Phi}_{\perp,t}, \beta_{t}, r_t, n_{\mathbf{H}}$\\$\gets \mathrm{adapt\_H}(\tilde{L}, \nabla \ln \tilde{L},q^{(r)}(\bar{\bm{u}}_r,\bm{v}_{r,t-1}),\bm{\Phi}_{r,t-1},\bm{\Phi}_{\perp,t-1}, \beta_{t-1}, \alpha_{\mathbf{H}})$
		\STATE Compute the required number of samples $n$ in function of $r_t$ with \cref{eq:n_r}
		\STATE Sample $\bar{\bm{\mathrm{U}}}_{r,\mathrm{add}} \in \mathbb{R}^{(n - n_{\mathbf{H}}) \times r_t} \gets \{\bar{\bm{u}}_{r,k} \stackrel{i.i.d.}{\sim} q^{(r)}(\bar{\bm{u}}_r,\bm{v}_{r,t-1})\}_{k = 1}^{n - n_{\mathbf{H}}} $
		\STATE Sample \smash{$\bar{\bm{\mathrm{U}}}_{\perp,\mathrm{add}} \in \mathbb{R}^{(n - n_{\mathbf{H}}) \times (d-r_t)} \gets \{\bar{\bm{u}}_{\perp,k} \stackrel{i.i.d.}{\sim} \varphi_{d-r_t}(\bar{\bm{u}}_\perp)\}_{k = 1}^{ n - n_{\mathbf{H}}}$}
		\STATE Compute $\bm{\ell}_{\mathrm{add}} \in \mathbb{R}^{(n - n_{\mathbf{H}}) \times 1} \gets \tilde{L}(\bar{\bm{U}}_{r,\mathrm{add}}^\tran \bm{\Phi}_{r,t}  + \bar{\bm{U}}_{\perp,\mathrm{add}}^\tran \bm{\Phi}_{\perp,t})$
		\STATE Join vertically $\bar{\bm{\mathrm{U}}}_r \gets [\bar{\bm{\mathrm{U}}}_{r,0},\bar{\bm{\mathrm{U}}}_{r,\mathrm{add}}],~\bar{\bm{\mathrm{U}}}_\perp \gets [\bar{\bm{\mathrm{U}}}_{\perp,0},\bar{\bm{\mathrm{U}}}_{\perp,\mathrm{add}}],~\bm{\ell} \gets [\bm{\ell}_{0},\bm{\ell}_{\mathrm{add}}]$

		\STATE Compute $\widehat{\bm{v}}_{t,\text{adj}}(\widehat{\bm{v}}_{r,t-1},\bm{\Phi}_{r,t-1},\bm{\Phi}_{r,t},\bm{\Phi}_{\perp,t-1},\bm{\Phi}_{\perp,t})$ according to \cref{eq:cotrafo}
		\STATE Compute the adjusted weights $\mathbf{w}_{t,\text{adj}} \in \mathbb{R}^{n \times 1} \gets w_{t,\text{adj}}(\bar{\bm{\mathrm{U}}}_{r},\bar{\bm{\mathrm{U}}}_{\perp},\bm{\ell},\widehat{\bm{v}}_{t,\text{adj}})$ with \cref{eq:adjusted_weights}
		\STATE Compute $\widehat{\bm{v}}_{r,t}$ with \cref{eq:CEBUred_parameter_estimate}
		\ENDWHILE
		\STATE Evaluate $\bm{\mathrm{w}}_{\mathrm{final}} \in \mathbb{R}^{n\times 1} \gets \left\{ \frac{ \tilde{L}(\bar{\mathbf{U}}_r^\tran \bm{\Phi}_{r,t}  + \bar{\mathbf{U}}_{\perp,t}^\tran \bm{\Phi}_\perp)\varphi_{r,t}(\bar{\mathbf{U}}_r^\tran \bm{\Phi}_{r,t})}{q^{(r)}(\bar{\mathbf{U}}_r^\tran \bm{\Phi}_{r,t},\widehat{\bm{v}}_{r,t})}\right\}_{k = 1}^n $
		\STATE Estimate evidence $\widehat{Z} \gets \frac{1}{n} \sum_{k=1}^n \mathrm{w}_{\mathrm{final},k}$
		\STATE Normalize weights $\bar{\bm{\mathrm{w}}}_{\mathrm{final}} \gets \bm{\mathrm{w}}_{\mathrm{final}}/(n \widehat{Z})$ 
		\STATE $\mathbf{U}_{\mathrm{post}} \gets$ Resample (with replacement) $N$ times from $\bar{\mathbf{U}}_r^\tran \bm{\Phi}_{r,t}  + \bar{\bm{U}}_{\perp,t}^\tran \bm{\Phi}_\perp$ with weighting $\bar{\bm{\mathrm{w}}}_{\mathrm{final}}$
		\STATE $\mathbf{\Theta}_{\mathrm{post}} \gets T^{-1}(\mathbf{U}_{\mathrm{post}})$ 
		\RETURN{$\mathbf{\Theta}_{\mathrm{post}}$, $\widehat{Z}$}
	\end{algorithmic}
\end{algorithm}
\section{Experimental results}
\label{sec:experiments}
We perform two numerical examples to demonstrate the capability and test for potential limitations of CEBUred. In the first example, we compare the computational cost and accuracy of CEBUred and CEBU in dependency of the ambient space dimension and verify the results with an analytical solution. In the second example, we examine the performance of CEBUred for different error thresholds as defined by \cref{eq:criterion}. Both methods are implemented with a Gaussian model as parametric IS density. In both examples, we infer a material parameter random field based on model output observations.
We measure the quality of posterior random field approximations $\widehat{Y}$ against a reference solution $Y$ (either analytical or numerical) in terms of the following spatially averaged relative mean and variance errors:
\begin{equation}
    \varepsilon_{\mu_Y} = \frac{\lVert \mu_Y(\bm{x}) - \hat{\mu}_Y(\bm{x}) \rVert_2} { \lVert \mu_Y(\bm{x}) \rVert_2} \quad \text{and} \quad 
    \varepsilon_{\sigma^2_Y} = \frac{\lVert \sigma^2_Y(\bm{x}) - \hat{\sigma}^2_Y(\bm{x}) \rVert_2} { \lVert \sigma^2_Y(\bm{x}) \rVert_2} . 
    \label{eq:relErrMeas}
\end{equation}
\subsection{1D Cantilever beam}
\label{subsec:beam}
\subsubsection{Problem description}
\label{subsubsec:beam_description}
\begin{figure}[!ht]
    \begin{minipage}[t]{0.46\textwidth}%
    	\centering%
    	\includegraphics{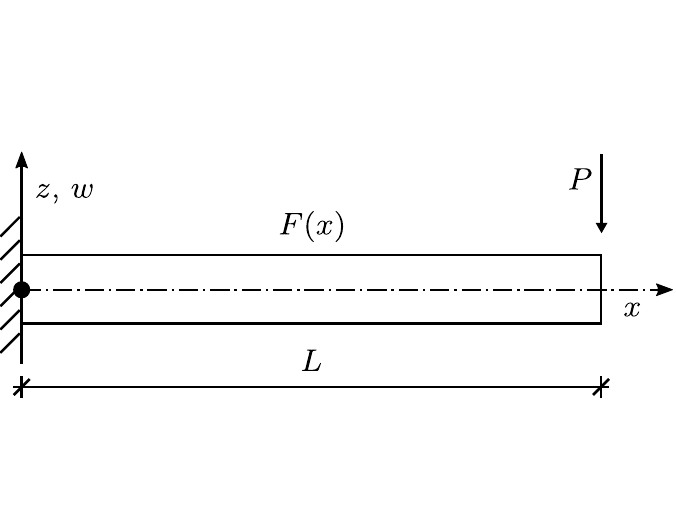}{}%
    	\caption{Cantilever beam with point load.}%
    	\label{fig:beam}%
    \end{minipage}%
    \hspace{5mm}%
    \begin{minipage}[t]{0.50\textwidth}%
    	\centering%
    	\includegraphics{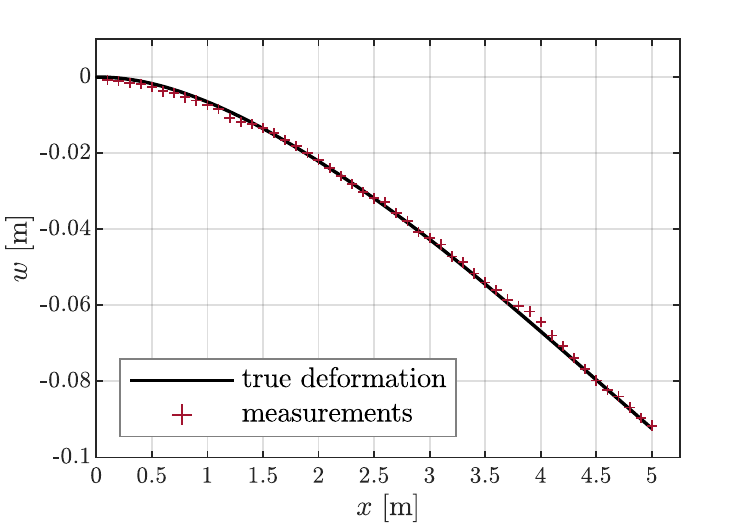}{}%
    	\caption{True deformation and measurements.}%
    	\label{fig:beam_defl}%
    \end{minipage}%
\end{figure}

We consider an Euler-Bernoulli beam with one clamped and one free end. It has length $L=5~\mathrm{m}$ and a point load of $P=20~\mathrm{kN}$ acting on the free end (\cref{fig:beam}). Its bending moment $M(x)$ can be obtained from the Euler-Bernoulli equation and reads \cite{bower2009applied}
\begin{equation}
M(x) = -E(x)I(x) \frac{d^2w(x)}{dx^2} ,
\label{eq:M_EB}
\end{equation}
where $E(x)$ is the beam's Young's modulus and $I(x)$ is its moment of inertia. Both can be summarized as the beam's axial flexibility $F(x)=1/E(x)I(x)$. The bending moment of the cantilever beam is computed as $M(x)=-P(L-x)$. Hence, the vertical deformation is given by 
\begin{equation}
w(x,F(x)) = P \int_0^x \int_0^s (L-x) F(x) \mathrm{d}t \mathrm{d}s  .
\label{eq:w_EB}
\end{equation}
The axial flexibility $F(x)$ is considered uncertain and spatially variable along the beam axis. We assign a homogeneous Gaussian prior random field with mean $\mu_F=10^{-4} \mathrm{kN}^{-1} \mathrm{m}^{-2} $, standard deviation $\sigma_F=3.5\cdot 10^{-5} \mathrm{kN}^{-1} \mathrm{m}^{-2} $ and exponential autocorrelation kernel
\begin{equation}
\rho(x,x'; l)=\exp{\left( - \frac{\lVert x-x'\rVert_1}{l} \right)}  ,
\label{eq:expMod}
\end{equation}
where $l$ is the correlation length. The correlation length of the random field of the axial flexibility is $l_F=2 \mathrm{m}$.
The forward model is given by a finite element (FE) model employing $100$ Euler-Bernoulli beam elements with cubic shape functions. The goal is to obtain samples of the posterior distribution of the axial flexibility given $n_{\mathrm{obs}} = 50$ equally spaced measurements $\{x_{\mathrm{meas},i}\}_{i=1}^{n_{\mathrm{obs}}}$ of the vertical deformation along the beam axis (see \cref{fig:beam_defl}).
Adjoint methods  \cite{Arora1979} are a computationally efficient tool for obtaining the model gradients required to compute $\nabla \tilde{L}$ as long as the number of model outputs of which derivatives are computed ($n_{\mathrm{obs}}$) is smaller than the number of model inputs ($d$) with respect to which derivatives are computed.
In the context of this example, the adjoint method is thus used for the setting $d=100$ only and the direct method is used in all other settings.
\\~\\
We assume the measurements to be corrupted by the additive, centered Gaussian noise vector $\bm{\eta} \sim \mathcal{N}(\bm{0},\bm{\Sigma}_{\eta\eta})$. The noise covariance matrix is defined as $[\bm{\Sigma}_{\eta\eta}]_{ij} = \sigma_\eta^2 \rho(x_{\mathrm{meas},i},x_{\mathrm{meas},j})$ with noise standard deviation $\sigma_\eta = 0.001 \mathrm{m}$, exponential correlation kernel $\rho(\cdot,\cdot;l_\eta)$ and correlation length $l_\eta=1 \mathrm{m}$. The random vector describing the vertical deformations in data space $\mathbb{R}^{n_{\mathrm{obs}}}$, i.e., at the $n_{\mathrm{obs}}$ measurement locations $\{x_{\mathrm{meas},i}\}_{i=1}^{n_{\mathrm{obs}}}$ is defined as
\begin{equation}
\tilde{\bm{w}} = \bm{w} + \bm{\eta}.
\label{eq:w_tilde}
\end{equation}
Given a set of realizations of $\tilde{\bm{w}}$, i.e., observational data $\tilde{\bm{y}}$, the likelihood function reads
\begin{equation}
L(\bm{F}) = \frac{1}{\sqrt{ (2\pi)^{n_{\mathrm{obs}}} \det( \bm{\Sigma}_{\eta\eta} ) } } \exp{ \left( [\tilde{\bm{y}} - \mathcal{G}(\bm{F})] \bm{\Sigma}_{\eta\eta}^{-1} [ \tilde{\bm{y}} - \mathcal{G}(\bm{F}) ]^\tran \right) }   ,
\label{eq:beamLike}
\end{equation}
where $\mathcal{G}(\cdot)$ represents the FE-model and returns the vertical deformations of the beam at the $n_{\mathrm{obs}}$ measurement locations.
\\~\\
The measurements for this example are obtained by generating a single random realization of the prior random field of the axial flexibility, solving \cref{eq:w_EB} numerically at $1001$ equally spaced discretization points and then adding randomly generated noise according to \cref{eq:w_tilde} to the solutions at the locations of the measurements $\{x_{\mathrm{meas},i}\}_{i=1}^{n_{\mathrm{obs}}}$. By using the analytical expression instead of the  FE-model for the generation of the measurements, we avoid the so-called 'inverse crime' \cite{Kaipio2005}.
\subsubsection{Analytical posterior} 
The following derivations closely follow \cite{Uribe2020b} where the example is investigated as well.
Since $F(x)$ is Gaussian and $w(x,F(x))$ is a linear function of $F(x)$ \cref{eq:M_EB}, the prior distribution of $w(x)$ is also Gaussian. Its mean and covariance read
\begin{subequations}
\begin{equation}
\mu_w(x) = P \int_0^x \int_0^s (L-t) F(t) \mathrm{d}t \mathrm{d}s = \frac{P \mu_F}{6} x^2 (3L-x) \quad \text{and}
\label{eq:mu_w}
\end{equation}
\begin{equation}
\Sigma_{ww}(x,x') = P \int_0^{x'} \int_0^x \int_0^{s'} \int_0^s (L-t) (L-t') \Sigma_{FF}(t,t') \mathrm{d}t \mathrm{d}t' \mathrm{d}s \mathrm{d}s'  .
\label{eq:Sig_ww}
\end{equation}
\end{subequations}
The explicit expression of \cref{eq:Sig_ww} is obtained using a computer algebra system and omitted here due to its tedious form.
\\~\\
An analytical solution of the posterior of the axial flexibility can be derived, since both the prior and the likelihood are Gaussian \cite{robert2007bayesian}. To this end, the Gaussian random vector $\bm{F}'= [\bm{F};\tilde{\bm{w}}]$ is considered, which contains the discretized random flexibility field, $\bm{F}\in \mathbf{R}^n$ and the  $n_{\mathrm{obs}}$ deformation measurements $\tilde{\bm{w}} \in \mathbf{R}^{n_{\mathrm{obs}}}$. The mean vector and covariance matrix of $\bm{F}'$ may be partitioned as 
\begin{equation}
\bm{\mu}_{F'} = 
\begin{bmatrix}
\bm{\mu}_F \\
\bm{\mu}_{\tilde{w}} 
\end{bmatrix}
\quad \text{and} \quad
\bm{\Sigma}_{F'F'} = 
\begin{bmatrix}
\bm{\Sigma}_{FF}            &\bm{\Sigma}_{F\tilde{w}} \\
\bm{\Sigma}_{F\tilde{w}}^\tran    &\bm{\Sigma}_{\tilde{w} \tilde{w}} 
\end{bmatrix}  .
\label{eq:F'part}
\end{equation}
As $\bm{F}'$ is jointly Gaussian, the posterior $\bm{F}|\tilde{\bm{y}}$ is Gaussian as well and has PDF
\begin{equation}
p(\bm{f}|\tilde{\bm{y}}) = \frac{1}{\sqrt{ (2\pi)^n \det( \bm{\Sigma}_{FF|\tilde{y}} ) }  } \exp{ \left( -\frac{1}{2} [\bm{f} - \bm{\mu}_{F|\tilde{y}}]^\tran \bm{\Sigma}_{FF|\tilde{y}}^{-1} [\bm{f} - \bm{\mu}_{F|\tilde{y}}] \right) }   .
\label{eq:postFlex}
\end{equation}
The posterior mean and covariance matrix are equal to the following conditional mean $\bm{\mu}_{F|\tilde{y}}$ and covariance matrix $\bm{\Sigma}_{FF|\tilde{y}}$:
\begin{equation}
\bm{\mu}_{F|\tilde{y}} = \bm{\mu}_F + \bm{\Sigma}_{F\tilde{w}} \bm{\Sigma}_{\tilde{w}\tilde{w}} ^{-1} (\tilde{\bm{y}} - \bm{\mu}_{\tilde{w}} )
\quad \text{and} \quad
\bm{\Sigma}_{FF|\tilde{y}} = \bm{\Sigma}_{FF} - \bm{\Sigma}_{F\tilde{w}} \bm{\Sigma}_{\tilde{w}\tilde{w}}^{-1} \bm{\Sigma}_{F\tilde{w}}^\tran
\label{eq:condPostFlex}  .
\end{equation}
All quantities in \cref{eq:condPostFlex} are computed within the partition in \cref{eq:F'part} except from $\bm{\mu}_{\tilde{w}}$, which is obtained by $\mathbb{E}[\tilde{\bm{w}}] = \mathbb{E}[\bm{w} + \bm{\eta} ] = \bm{\mu}_w $.   
\subsubsection{Parameters of numerical study}
The flexibility random field is discretized in space using a midpoint method \cite{DerKiureghian1988} with $d$ collocation points. $d$ is therefore the ambient space dimension of the Bayesian inverse problem, where scenarios $d = \{10, 25, 50, 100\}$ are investigated.
We use CEBU and CEBUred to obtain samples from the $d$-dimensional posterior distribution of the axial flexibility given a set of $n_{\mathrm{obs}} = 50$ measurements.
We use $\delta_{v,target}=1.5$ and $n = 1000$ samples per level for all CEBU runs. 
For CEBUred we chose $\delta_{v,target}=1.5$, $\alpha_{\mathrm{par}} =4$, $\alpha_{\mathbf{H}}=6$ and $\varepsilon=1.0$. Results are averaged over $54$ repeated runs of both CEBU and CEBUred.
\subsubsection{Discussion of results}
\begin{figure}[!ht]
    \centering
    \includegraphics{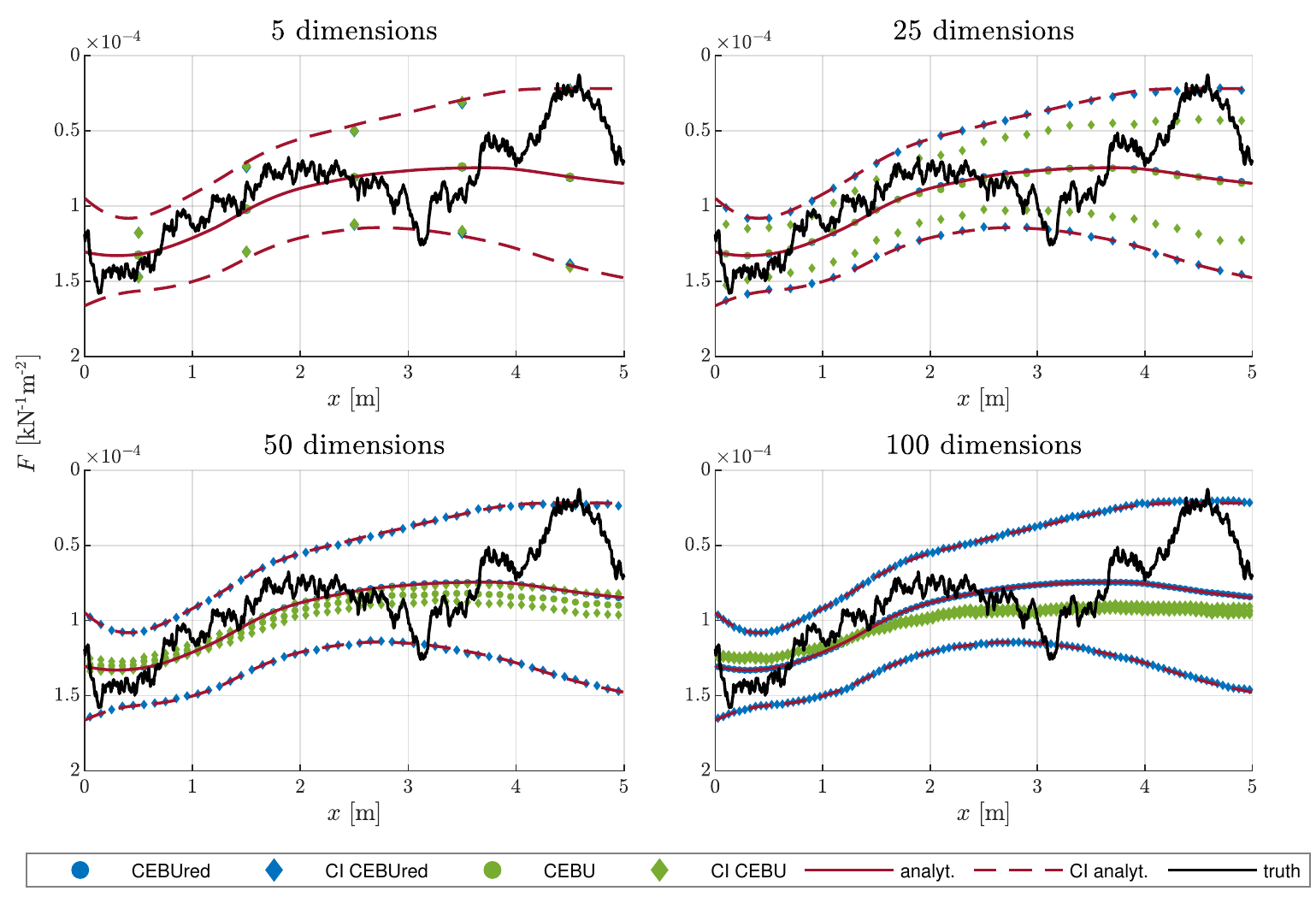}{}
    \caption{Axial flexibility posterior field: mean and 95\% credible intervals. CEBU and CEBUred solutions are plotted at the random field collocation points (midpoints).}
	\label{fig:beamPostFields}    
\end{figure}

\cref{fig:beamPostFields} shows the posterior flexibility fields obtained with both CEBU and CEBUred at varying ambient dimension. At $d=5$, the results obtained with both CEBU and CEBUred coincide with the analytical reference posterior as indicated by the almost congruent scatter points in the top left panel of \cref{fig:beamPostFields}. However, discretizing the flexibility field with only $5$ subparts does not allow for an accurate representation of the posterior field at the clamping. There, the axial flexibility exerts the strongest influence on the beam deformation thus requiring a finer discretization.
\\~\\
As $d$ increases, the results obtained by CEBU deteriorate, as indicated by both an increasing deviation of the CEBU solution from its analytical counterpart in both mean and 95\% posterior credible bounds. At $d > 11$, CEBU has too few samples available to accurately fit all biasing density parameters in ambient space. CEBUred, on the other hand, agrees closely with the analytical solution if the chosen discretization is fine enough.
\begin{figure}[!ht]
\centering
    \begin{minipage}[t][][t]{0.59\textwidth}
        \vspace{0pt}%
    	\centering%
    	\includegraphics{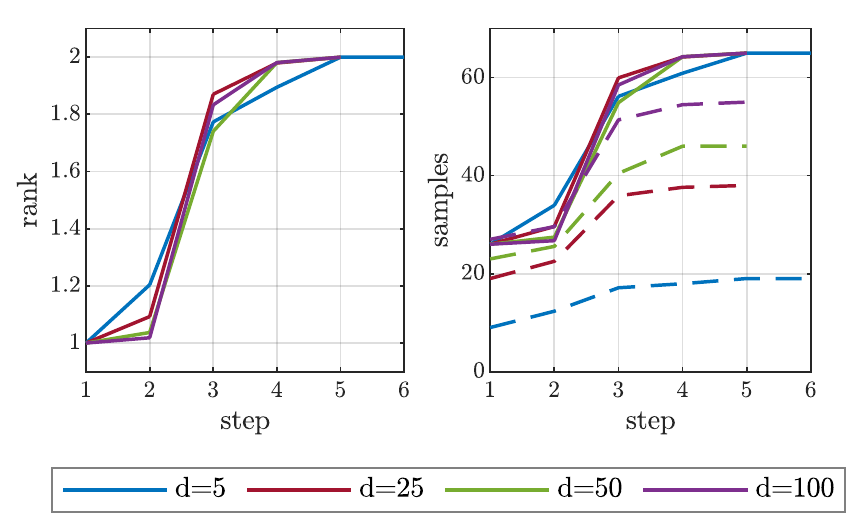}%
    	\caption{Left: Average number of selected LIS dimension with increasing CEBUred step index. Right: Average number of evaluated samples per step with increasing CEBUred step index. The solid and the dashed lines represent the total number of model and model gradient evaluations, respectively.}
    	\label{fig:beamRanksSamples}
    \end{minipage}%
    \hspace{3mm}%
    \begin{minipage}[t][][t]{0.39\textwidth}
        \vspace{0pt}%
    	\centering%
    	\includegraphics{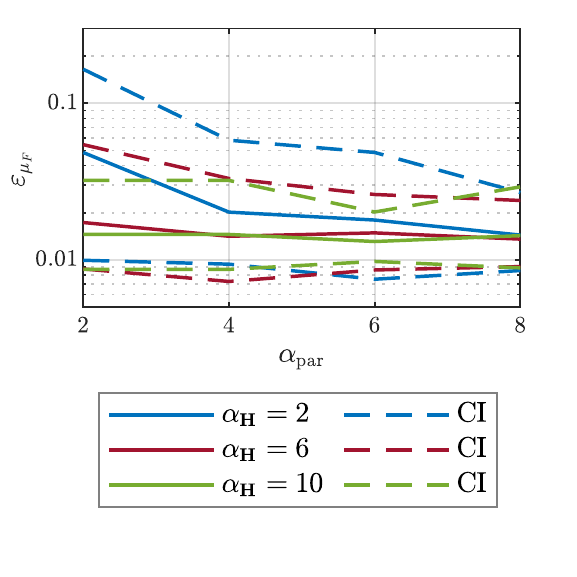}%
    	\caption{Relative posterior mean error for different combinations of $\alpha_{\mathbf{H}}$ and $\alpha_{\mathrm{par}} $ at $d = 100$.}
    	\label{fig:beamFudgePars}
    \end{minipage}%
\end{figure}
\\~\\
\cref{fig:beamRanksSamples} shows the number of selected ranks $r$ (corresponds to the dimension of the effectively used subspace (LIS) in CEBUred) and number of samples plotted over the CEBUred step index. The number of LIS dimensions reduces to $r=1$ within the first step for all tested $d$. The number of runs per number of steps for different $d$ are shown in \cref{tab:beamRunsSteps}. At $d=5$, one of the $54$ runs terminated after $6$ steps, whereas all other simulations terminated after a maximum of $5$ steps. \cref{tab:beamRunsSteps} suggests that for the given FE-discretization, finer random field discretizations tend to stabilize the simulation in the sense that most runs require the same number of steps. 
%
%

\begin{table}[!ht]
	\label{tab:beamRunsSteps}
	\centering
	\caption{Number of runs broken down according to required number of CEBUred steps at varying $d$.}
	\small{
		\begin{tabular}{ccccc}
			\hline
			\# of steps & $d=5$ &  $d=25$ & $d=50$ & $d=100$\\ 
			\hline
			2 & $1$ & $0$ & $0$ & $0$\\
			3 & $15$ & $5$ & $1$ & $1$\\
			4 & $29$ & $43$ & $42$ & $46$\\
			5 & $8$ & $6$ & $11$ & $7$\\
			6 & $1$ & $0$ & $0$ & $0$\\
			\hline                        
		\end{tabular}
	}
\end{table}
~\\
\begin{table}[]
	\centering
	\caption{Likelihood and gradient evaluations for the beam example per run of CEBU and CEBUred (CEBUred: averaged over 54 runs).}
	\label{tab:beam_cost}
	\small{
		\begin{tabular}{lccc}
			\hline
			Problem  & CEBU&   CEBUred  & CEBUred\\ 
			dimension & (Likelihood calls) &    (Likelihood calls)  &  (Gradient calls) \\ 
			\hline
			\hline
			 $d=5$    & 3796 & 170.0 &  54.3 \\ 
			\hline                     
			$d=10$ & 3870  & 181.0 &  115.8 \\
			\hline
			 $d=25$     & 3833 & 184.6 &  143.5\\ 
			\hline
			 $d=100$     & 3833  &182.7  & 168.6\\ 
			\hline
		\end{tabular}
	}
\end{table}
For all investigated ambient dimensions $d$ and in each CEBUred step, the beam problem possesses very low-dimensional (likelihood-informed) subspaces, in which the inverse problem can be solved efficiently ($r = 1-2$). In this LIS, significantly less samples are required to accurately characterize biasing densities compared to CEBU, which operates in $d$-dimensional ambient space. 
\cref{fig:beamRanksSamples} (solid lines) illustrates the correspondence of the number of required samples in CEBUred with the reduced space dimension (rank $r$). As $d$ increases, more gradients are evaluated (\cref{fig:beamRanksSamples}, right, dashed lines). This is due to the factor $\ln d$ in \cref{eq:n_H}.
\\~\\
\begin{table}[!ht]
	\label{tab:beamRelErr}
	\centering
	\caption{Relative posterior mean and variance error at varying $d$ averaged over 54 repeated runs of CEBUred.}
	\small{
		\begin{tabular}{ccccc}
			\hline
			error & $d=5$ &  $d=25$ & $d=50$ & $d=100$\\ 
			\hline
			$\varepsilon_{\mu_F}$ & $0.0588$ & $0.0219$ & $0.0167$ & $0.0142$ \\
			$\varepsilon_{\sigma^2_F}$ & $0.2168$ & $0.1039$ & $0.1090$ & $0.0907$\\
			\hline                        
		\end{tabular}
	}
\end{table}
In \cref{tab:beam_cost} we list the average number of required likelihood and likelihood gradient evaluations for both CEBU and CEBUred. The number of required likelihood evaluations within CEBUred remains approximately constant across all investigated dimensions and is more than an order of magnitude lower compared to number of evaluations required by CEBUred. The numbe of likelihood gradient evaluations grows with $d$ but remains below the number of likelihood evaluations. Depending on the method of evaluating these gradients, a gradient call may however be considerably more expensive than a likelihood call.
\\~\\
$\mu_F(\bm{x})$ and $\sigma^2(\bm{x})$ are the analytical posterior mean and variance, respectively, evaluated at the discretization points and $\hat{\mu}_F(\bm{x})$ and $\hat{\sigma}^2_F(\bm{x})$ are their sample-based counterparts obtained with CEBUred. \cref{tab:beamRelErr} shows the relative error of the mean and the variance for the different dimensions $d$. What is not immediately obvious from the plots in \cref{fig:beamPostFields} is that a finer discretization indeed leads to smaller relative errors. However, the decrease slows down from $d=25$ and is rather small between $d=50$ and $d=100$. 
\cref{fig:beamFudgePars} shows the relative posterior mean error 
$\varepsilon_{\mu_F}$ for different combinations of the fudge factors $\alpha_{H}$ and $\alpha_{\mathrm{par}} $ at $d=100$. 
 The relative posterior mean error decreases signifcantly between $\alpha_{\mathbf{H}}=2$ and $\alpha_{\mathbf{H}}=6$ and remains constant as $\alpha_{\mathbf{H}}$ is increased from $6$ to $10$.
This is likely due to the fixed error threshold of $\epsilon=1.0$, which prescribes the approximation quality of the optimal projector. Once the relevant part of the $\mathbf{H}$-spectrum (corresponding to the choice of $\epsilon$) is estimated accurately, increasing $\alpha_{\mathbf{H}}$ bears no further effect. In this case, larger values of $\alpha_{\mathrm{par}}$ only lead to a larger number of model evaluations, which in turn can lead to a better fit of the biasing density.
At $\alpha_{\mathbf{H}}=2$, the error decreases with increasing $\alpha_{\mathrm{par}}$, whereas it remains approximately constant when increasing $\alpha_{\mathrm{par}}$ at $\alpha_{\mathbf{H}}\ge 6$.
In the latter case, the large number of gradient evaluations (each of which also yields a model evaluation) are already sufficient to accurately estimate the parameters of the reduced biasing density such that increasing $\alpha_{\mathrm{par}}$ will not further reduce the error.
\subsection{2D plate in plane stress}
The example was first presented in \cite{Liu1993} in the context of uncertainty quantification. We consider the adapted version from \cite{Uribe2020b}. Through this example, we investigate how the accuracy of the resulting posterior improves by using different error thresholds $\epsilon$ as defined by \cref{eq:criterion}. 
\subsubsection{Problem description}
\begin{figure}[!ht]
    \centering
    \includegraphics{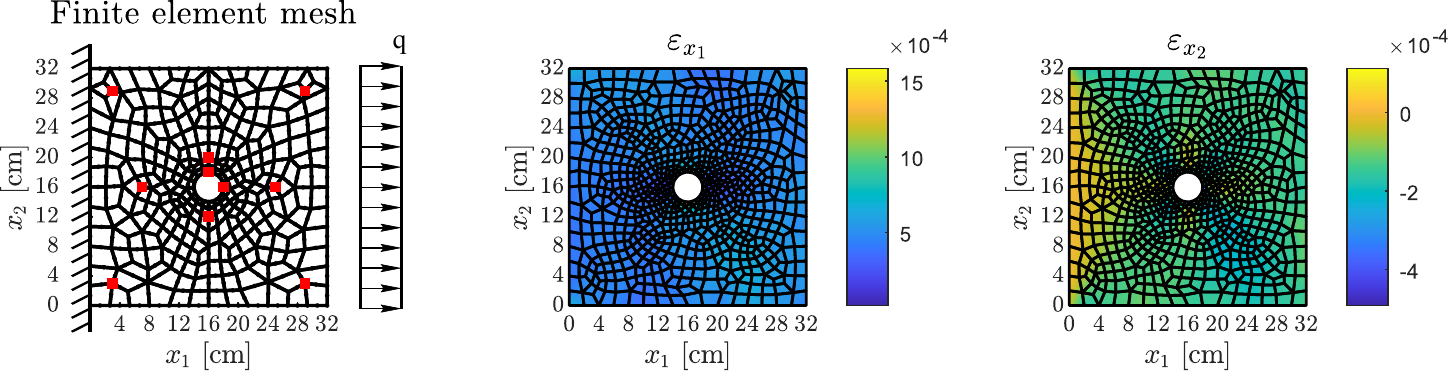}%
    \caption{Left: FE-model of the plate. The red squares mark the positions of the strain gauges. Center and right: True fields of the strains in $x_1$- and $x_2$-direction.}
    \label{fig:plateTrueStrains}
\end{figure}

We consider a 2D square steel plate in plane-stress with side length $32~\mathrm{cm}$, thickness $t=1~\mathrm{cm}$ and a hole with radius $r=2~\mathrm{cm}$ located at its center (\cref{fig:plateTrueStrains}). The plate is clamped at the left-hand side and loaded with a constant line load $q=6~\mathrm{kN}~\mathrm{cm}^{-2}$ acting on its right-hand side. The plate has density $\rho=7850~\mathrm{kg}~\mathrm{m}^{-3}$, which is required to account for body forces (oriented in negative $x_2$-direction), and the Poisson ratio is $\nu=0.29$.
%

Assuming plane stress, the displacement field  $\bm{u}(x_1,x_2)=[u_{x_1}(x_1,x_2),u_{x_2}(x_1,x_2)]^\tran$ can be computed implicitly based on elasticity theory through a set of elliptic PDEs (Cauchy-Navier equations) \cite{Johnson2012}:
\begin{equation}
    G(x_1,x_2)\nabla^2\bm{u}(x_1,x_2)+\frac{E(x_1,x_2)}{2(1-\nu)}\nabla (\nabla \cdot \bm{u}(x_1,x_2))+\bm{B} = 0  .
    \label{eq:ellPDE}
\end{equation}
$G(x_1,x_2):=E(x_1,x_2)/(2(1+\nu))$ is the shear modulus, $E(x_1,x_2)$ is Young's modulus, and $\bm{B}=[b(x_1),b(x_2)]^\tran$ is the vector of body forces acting on the plate. In order to solve \cref{eq:ellPDE}, an FE model with $282$ eight-noded quadrilateral finite elements is used (\cref{fig:plateTrueStrains}).
\\~\\
In this example, the plate's Young's modulus is considered uncertain and spatially variable. We assign a homogeneous random field prior with log-normal marginal distributions with mean $\mu_{E}=2\cdot 10^4 ~\mathrm{kN}~\mathrm{cm}^{-2}$ and standard deviation $\sigma_{E}=3 \cdot 10^3 ~\mathrm{kN}~\mathrm{cm}^{-2}$. The mean and standard deviation of the underlying Gaussian field $\ln E(x_1,x_2)$ follow as $\mu_{\ln E}=9.89$ and $\sigma_{\ln E}=0.15$, respectively, and its correlation structure is modelled with the exponential kernel of \cref{eq:expMod} and correlation length $l_{\ln E} = 10~ \mathrm{cm}$.
\\~\\
We discretize $\ln E$ by means of a Karhunen- Lo{\`e}ve-expansion (KL-expansion). To this end, we solve the following homogeneous Fredholm integral equation of the second kind \cite{ghanem1991}
\begin{equation}
    \sigma_{\ln E}^2 \int_D \rho(\bm{x}, \bm{x}'; l_{\ln E}) \phi_k(\bm{x}') \mathrm{d} \bm{x}' = \lambda_k \phi_k(\bm{x}) 
    \label{eq:Fred2}
\end{equation}
for the covariance kernel's set of eigenpairs $\{\lambda_k,\phi_k\}$. Consequently, we can express the log-normal Young's modulus prior as the $\exp$ of a KL-expansion \cite{ghanem1991} like
\begin{equation}
    E(x_1,x_2;\bm{\theta}) = \exp{ \left[ \mu_E + \sum_{k=1}^\infty \sqrt{\lambda_k} \phi_k(x_1,x_2) \theta_k \right] }  ,
    \label{eq:E_KL}
\end{equation}
where the coefficients $\theta_k$ are pairwise independent standard-normal Gaussian random variables.
\\~\\
We estimate the set of eigenpairs $\{\lambda_k,\phi_k\}$ for the KL-expansion by solving \cref{eq:Fred2} using the Nystr{\"o}m method on a grid of $160\times160$ Gauss-Legendre quadrature points. The eigenfunctions are interpolated at the numerical integration points of the elements of the FE-model \cite{press1986numerical}.
\\~\\
Truncating the KL-expansion \cref{eq:E_KL} after $M$ terms results in an $M$-order KL-approximation of $E$, which we denote as $\hat{E}(\bm{x};\bm{\theta})$. 
This approximation recovers the random field mean exactly, however is associated with an under-representation of its variance $ \ln \sigma_E^2$. This under-representation is often measured with the global relative variance error of the $M$-order KL-approximation:
\begin{equation}
\label{eq:globErrField}
   \bar{\epsilon}_{\ln \sigma^2} = \frac{1}{|D|} \int_D  \left| \frac{ \mathbb{V}[E(\bm{x};\bm{\theta})] - \mathbb{V}[\hat{E}(\bm{x};\bm{\theta})] } { \mathbb{V}[E(\bm{x};\bm{\theta})] } \right|  \mathrm{d}\bm{x}  = 1 - \frac{1}{|D|\cdot \ln \sigma_E^2}\sum_{k=1}^M \lambda_k.
\end{equation}
Therein, $D$ is the spatial domain of the random field $E$.
The inference task for the example consists in learning the Young's modulus' posterior distribution based on strain measurements at $n_{\mathrm{obs}}=10$ positions on the plate. At each position, two gauges measure the strain in $x_1$- and $x_2$-direction (\cref{fig:plateTrueStrains}, left, red squares), respectively. Hence, a set of $20$ measurements are available to solve the inference task. We generate the measurements artificially by using a FE-model on a finer mesh of 779 elements, in order to once again avoid the 'inverse crime' \cite{Kaipio2005}. The true strains are depicted in the center and right plot in \cref{fig:plateTrueStrains}. 
\\~\\
The strain measurements are generated by solving the forward problem based on a single Young's modulus prior random field realization. This realization is generated using a midpoint method discretized at the numerical integration (Gauss) points of the plate FE model rather than a KL-approximation. Consequently, noise is added to the computed strains at the measurement locations. We model the noise as a centered Gaussian random vector $\bm{\eta} \sim \mathcal{N}(\bm{0},\bm{\Sigma}_{\eta\eta})$. The noise standard deviation is set to $\sigma_\eta=10^{-4}$ and the autocorrelation of both $x_1$- and $x_2$-strain measurements is modelled with the exponential kernel \cref{eq:expMod} using a correlation length of $l_\eta=10 ~\mathrm{cm}$. The cross-correlation function between $x_1$- and $x_2$-strain measurements is taken as the autocorrelation function multiplied by a cross-correlation coefficient of $0.25$.
\subsubsection{Numerical reference posterior}
We use \textit{adaptive Bayesian Updating with Subset Simulation} (aBUS-SuS) to verify the solution obtained with CEBUred. aBUS-SuS has been tested on a variety of engineering applications, e.g., in \cite{Straub2015a,Straub2016,Betz2018,Jiang2018}. aBUS-SuS recasts the Bayesian inverse problem as a structural reliability problem \cite{Uribe2020b}. \textit{Subset simulation} (SuS) \cite{Beck2001} is a robust and efficient method for solving such structural reliability problems and within aBUS-SuS, SuS is employed to solve general Bayesian inverse problems. SuS itself requires carrying out an MCMC sampling step for which we use a pCN sampler \cite{Cotter2013} with adaptive scaling \cite{Papaioannou2015MCMCAF}.

\subsubsection{Parameters of numerical study}
We discretize the Young's modulus random field by a KL-approximation \cref{eq:E_KL} with 879 terms producing a Bayesian inverse problem with ambient dimension $d=879$. The chosen number of terms accounts for at least $97\%$ of the spatial variance of the random field meaning the average variance error \cref{eq:globErrField} is  $\le 3\%$. 
\\~\\
The inference task is solved by using CEBUred with different error tolerances $\epsilon$. We choose $\epsilon=\{1.0,10^{-1},10^{-2},10^{-3}\}$. The remaining parameters are set as $\delta_{v,target}=1.5$, $\alpha_{\mathrm{par}} =3$, and $\alpha_{\mathbf{H}}=4$. The gradients of the likelihood function required at each step of CEBUred are evaluated with the adjoint method \cite{Arora1979} (derived for this particular problem in \cite[Appendix A]{Uribe2020a}). In the final step, we draw $N=10000$ samples from the approximate posterior distribution. The parameters for the numerical reference posterior generated with aBUS-SuS are $n=20000$ for both samples per subset level and final samples of the approximated posterior and intermediate conditional probability $p=0.1$.
Except for the reference posterior, which is computed once only, we repeat each analysis with CEBUred $40$ times and average all results over the individual runs.
\subsubsection{Discussion of results}
\begin{figure}[!ht]
    \centering
    \includegraphics{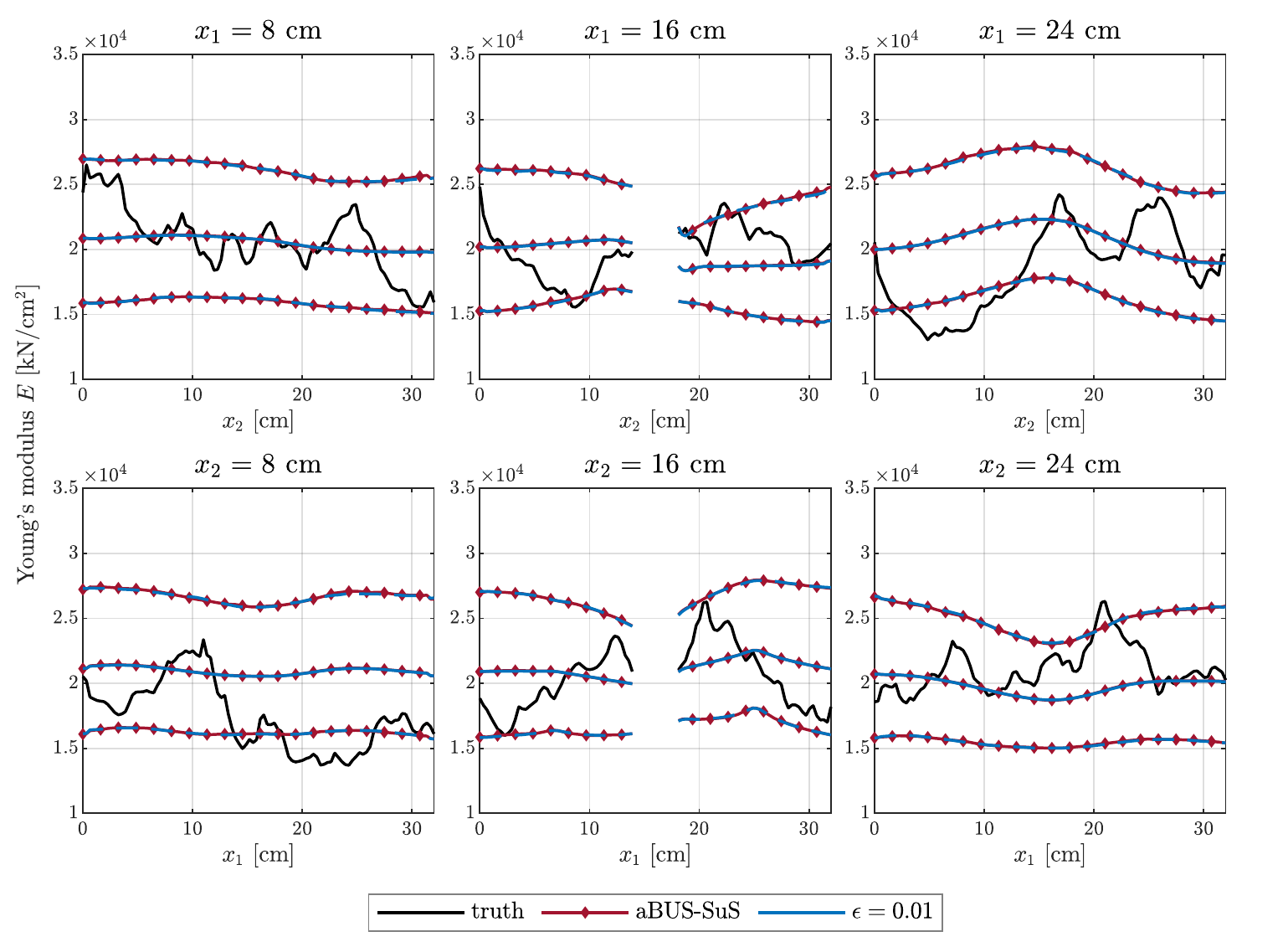}{}
    \caption{Posterior Young's modulus random field obtained with aBUS-SuS and with CEBUred using $\epsilon=10^{-2}$: Means (solid lines) and 95\% credible intervals (CI, dashed lines) at three vertical (top row) and three horizontal (bottom row) cross-sections.}
	\label{fig:platePostFields}    
\end{figure}

\begin{figure}[!ht]
\centering
    \begin{minipage}[t][][b]{0.65\textwidth}
        \vspace{0pt}%
    	\centering%
    	\includegraphics{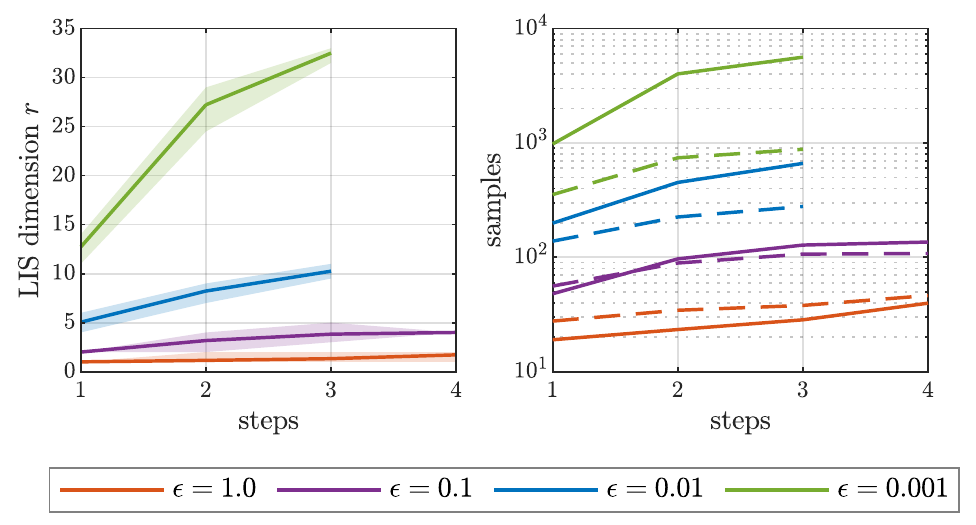}%
    	\caption{Left: Average number of selected LIS dimension with increasing CEBUred step index (95\% CI indicated as shaded area). Right: Average number of evaluated samples per step with increasing CEBUred step index. The \textbf{solid} and the \textbf{dashed} lines represent the total number of \textbf{model} and \textbf{model gradient} evaluations, respectively.}
    	\label{fig:plateRanksSamples}
    \end{minipage}%
    \hspace{5mm}%
    \begin{minipage}[t][][b]{0.30\textwidth}
        \vspace{0pt}%
    	\centering%
    	\includegraphics{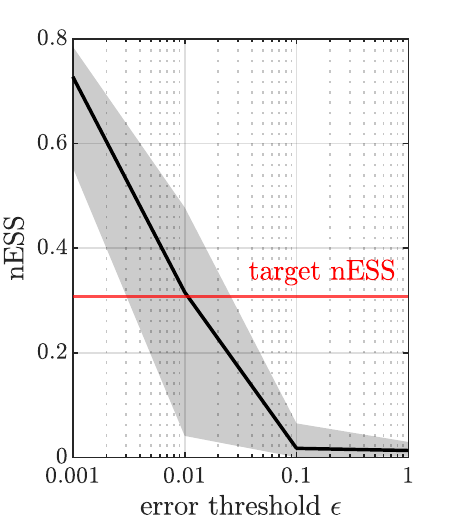}%
    	\caption{Mean of the normalized effective sample size (nESS)  (95\% CI indicated as shaded area) plotted over the tested error thresholds along with the target nESS (solid red line).}
    	\label{fig:plateNESS}
    \end{minipage}%
\end{figure}

\cref{fig:platePostFields} compares the posterior fields obtained with CEBUred using an error threshold of $\epsilon=10^{-2}$ and the reference posterior obtained with aBUS-SuS along six sections across the plate. Along each section, the CEBUred-based posterior means are in good agreement with the reference posterior mean. For all other tested error thresholds ($\epsilon \in \{1,10^{-1},10^{-2}\}$), similar results are obtained for the average posterior means and variances taken over 40 repeated CEBUred runs (see \cref{tab:plateRelErr}). The coefficients of variation of the posterior mean and variance estimates are rather large for $\epsilon \ge 10^{-1}$, but decrease significantly between $\epsilon = 10^{-1}$ and $\epsilon = 10^{-2}$ based on the results given in \cref{tab:plateRelErr}.
\\~\\
The numbers of required likelihood and likelihood gradient evaluations for aBUS-SuS and CEBUred at the four tested error threshold are listed in \cref{tab:plate_cost}. At $\epsilon = 10^{-2}$, CEBUred reduces the number of required likelihood calls by roughly two orders of magnitude compared to the reference aBUS-SuS run, which comes at the cost of 642 additional gradient calls. 
\\~\\
%
\cref{fig:plateRanksSamples} shows the mean ranks and the corresponding number of model and gradient evaluations. 
To ensure an accurate construction of the LIS, the eigenvectors corresponding to the $r$ largest eigenvalues of $\mathbf{H}$ must be reasonably well estimated. The number of samples required for the estimation is determined with the heuristic formula given by \cref{eq:n_H}. According to this formula, the number of samples for the estimation of $\mathbf{H}$ linearly depends on $r$. Therefore, the lines in the left plot and the dashed lines in the right plot in \cref{fig:plateRanksSamples} are linearly dependent. For $\epsilon \ge 10^{-1}$, we obtain LIS dimensions of $r\le5$ in all steps. 
\begin{table}[]
	\centering
	\caption{Likelihood and gradient evaluations for the plate example per run of aBUS-SuS and CEBUred (CEBUred: averaged over 40 runs).}
	\label{tab:plate_cost}
	\small{
		\begin{tabular}{lccccc}
			\hline
			 &aBUS-SUS  & CEBUred  &   CEBUred  & CEBUred  & CEBUred\\ 
			 &  &  ($\epsilon=1$) &    ($\epsilon=10^{-1}$)  &  ($\epsilon=10^{-2}$) &  ($\epsilon=10^{-3}$)\\ 
			\hline
			\hline
			Llikelihood calls  & 120000 & 76 & 276 & 1313 & 10621\\ 
			\hline
			Gradient calls     & - & 105.3 & 254.5 & 642.0 & 1975.1\\
			\hline
		\end{tabular}
	}
\end{table}

\begin{table}[!ht]
	\label{tab:plateRelErr}
	\centering
	\caption{Relative posterior mean and variance error at varying $\epsilon$ averaged over 40 runs with coefficient of variation in round brackets}
	\small{
		\begin{tabular}{ccccc}
			\hline
			error & $\epsilon=1$ &  $\epsilon=10^{-1}$ & $\epsilon=10^{-2}$ & $\epsilon=10^{-3}$\\ 
			\hline
			\rule{0pt}{3.0ex}   $\varepsilon_{\mu_E}$ (CoV)  & $0.017 (1.66)$ & $0.019 (0.98)$ & $0.003(0.32)$ & $0.002(0.17)$ \\
			\rule{0pt}{3.0ex} $\varepsilon_{\sigma^2_E}$ (CoV) & $0.174(0.52)$ & $0.201 (0.75)$ & $0.039(0.33)$ & $0.027(0.10)$\\		
			\hline                        
		\end{tabular}
	}
\end{table}
\cref{fig:plateRanksSamples} (right) depicts the number of gradient and model evaluations per CEBUred step for varying $\epsilon$ exposing that for $\epsilon = 1$, the number of gradient evaluations performed to estimate the LIS exceeds the overall number model evaluations required to perform the parameter update of the biasing density.  
This is due to the fact that the number of required model evaluations quadratically depends on the LIS dimension $r$ through \cref{eq:n_r}, where in turn $r$ decreases with increasing $\epsilon$.
\\~\\
\cref{fig:plateNESS} displays the nESS of the final posterior sample along with the target nESS, $n_{\mathrm{eff}}^*=0.3077$, that is related to the target coefficient of variation of the IS weights $\delta_w=1.5$ through \cref{eq:nESS}. The simulations with $\epsilon=1.0$ and $\epsilon=10^{-1}$ exhibit small nESS well below the target. In these cases, the dimensionality of the LIS is too low to accurately represent the posterior of the random field and, consequently, the IS weights have large variance. This leads to larger posterior mean and variance errors and larger associated coefficients of variation of these error measures for $\epsilon=1.0$ and $\epsilon=10^{-1}$ as documented in \cref{tab:plateRelErr}. 
\\~\\
Compared to the beam application, where $n_{\mathrm{eff}}^*$ was achieved for all investigated ambient dimensions with $\epsilon=1.0$, the plate obviously requires a stricter error threshold. According to \cref{tab:plateRelErr} and \cref{fig:plateNESS}, $\epsilon=10^{-2}$ is a good choice for the present example. However, if the threshold is chosen very small, e.g., $\epsilon=10^{-3}$, no significant improvement is observed. In this case, the marginally increased accuracy will not justify the additional computational expenses incurred by reducing the threshold.
\\~\\
\begin{table}[!ht]
	 \label{tab:plateRunsSteps}
	\centering
	\caption{Number of runs broken down according to required number of CEBUred steps at varying $\epsilon$.}
	\label{tab:plate_results}
	\small{
		\begin{tabular}{ccccc}
			\hline
			\# of steps & $\epsilon=1$ &  $\epsilon=10^{-1}$ & $\epsilon=10^{-2}$ & $\epsilon=10^{-3}$\\ 
			\hline
			2 & $3$ & $0$ & $0$ & $0$ \\
			3 & $30$ & $39$ & $40$ & $40$\\
			4 & $7$ & $1$ & $0$ & $0$\\
			\hline                        
		\end{tabular}
	}
\end{table}
\begin{figure}[!ht]
    \centering
    \includegraphics{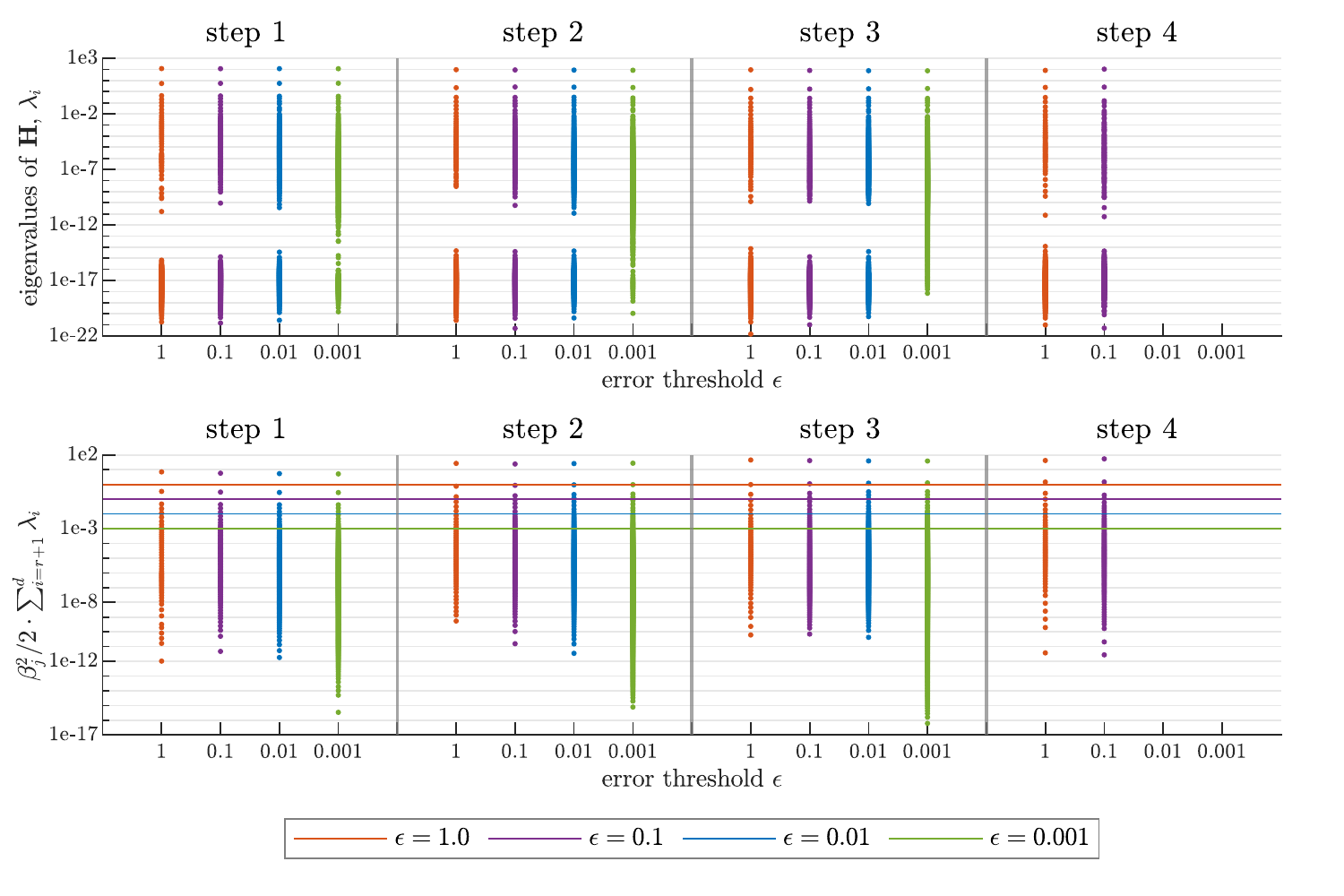}{}
    \caption{Top: $\mathbf{H}$-spectra plotted versus CEBUred step number for each error threshold $\epsilon$. Bottom: Upper bound of the KLD between the tempered, full posterior and the tempered, optimally reduced posterior as defined by \cref{eq:criterion}. For each $\epsilon$ and CEBUred step, the largest scatter point corresponds to $r=1$, the second-largest corresponds to $r=2$ and so on to $r=d-1$. Solid lines indicate the four tested values of the error threshold $\epsilon$.}
	\label{fig:plateScatter}    
\end{figure}
\cref{fig:plateScatter} (top) depicts the spectrum of $\mathbf{H}$ at each step of CEBUred and each predefined error threshold. As shown in \cref{tab:plateRunsSteps}, all runs with $\epsilon \le 10^{-2}$ required three steps, whereas only few runs with $\epsilon \ge 10^{-1}$ required four steps. For this reason, no spectra appear in the rightmost panels of \cref{fig:plateScatter} at step 4 for $\epsilon \le 10^{-2}$. All displayed spectra share two dominant eigenvalues of comparable magnitude followed by a sharp decay. The main difference amongst spectra associated with different $\epsilon$ is the number of samples used to estimate $\mathbf{H}$, $n_{\mathbf{H}}$. While $n_{\mathbf{H}}$ has no influence on the dominant eigenvalues, it bears some effect on a gap in the center of the spectrum, that becomes narrower and eventually closes as $n_{\mathbf{H}}$ increases. This effect, however, is negligible for the LIS construction as it takes place at eigenvalue magnitudes well below that of the smallest eigenvalue whose corresponding eigenvector is included in the LIS for any choice of $\epsilon$.
\\~\\
In \cref{fig:plateScatter} (bottom), we show the criterion \cref{eq:criterion} for each possible choice of $r=1 \dots d-1$ along with the tested error thresholds $\epsilon = \{1.0,10^{-1},10^{-2},10^{-3}\}$. At any given step $t$, we can directly compare the scatter points belonging to different error thresholds since they have equal $\beta_t$ on average. The number of scatter points that lie above the horizontal solid lines indicating $\epsilon$ determine the LIS dimension $r$. The upper scatter points are approximately equal for any choice of $\epsilon$ and any $t$  as the corresponding summations are dominated by their leading term. 
\\~\\
Using the heuristic given in \cref{eq:n_H} leads to a good estimate of the desired first eigenvalues and vectors. This is important because the accuracy of the LIS depends on these eigenvectors. However, for the computation of the upper bound of the KLD between the full posterior and the optimal reduced posterior \cref{eq:criterion} and the subsequent determination of the rank, all eigenvalues are needed. For problems with rapidly decaying eigenvalue spectra of their $\mathbf{H}$ matrix, as we see in this example, this is not a concern in practice, since their smallest eigenvalues have little effect on the computation of \cref{eq:criterion}.
\section{Concluding remarks}
\label{sec:conclusions}
We present  CEBUred (\textbf{C}ross-\textbf{E}ntropy-based IS method for \textbf{B}ayesian \textbf{U}pdating in \textbf{red}uced space), an algorithm for approximating posterior distributions that are the solutions of nonlinear Bayesian inverse problems. Such problems often arise in the context of finding inverse solutions to computationally expensive numerical models and solvers. Thus, computational efficiency is of the essence, which translates to minimizing the number of required samples (evaluations of the numerical model) to approximate the sought posterior distribution at a prescribed accuracy. We address high-dimensional problem settings that arise, e.g., if the inference target is represented by random fields or processes, by identifying low-dimensional linear subspaces \cite{Zahm2021} in which we perform cross-entropy-based importance sampling \cite{Engel2021}. These subspaces are obtained as truncated eigenspaces of the second-moment matrix of the gradient of the log-likelihood $\mathbf{H}$. 
\\~\\
We investigate CEBUred using two benchmark problems from engineering mechanics. In the first example, the material parameter random field of a cantilever beam subject to a point load is inferred from noise-distorted deflection measurements. We examine the performance of CEBU versus CEBUred versus a known analytical posterior reference solution at varying dimension of the material parameter random field discretization. We find that the dimensionality reduction is vital to ensure the posterior approximation accuracy is independent of the problem dimension by comparing CEBU and CEBUred. The second problem consists of inferring the material parameter random field of a clamped steel plate under load from a strain measurement at 10 locations on the plate. The inference problem is set in a $879$-dimensional space as the parameter random field is discretized with a $879$-term KLE. We find that the quality of the posterior approximation produced by CEBUred is closely connected to the choice of the error threshold $\epsilon$ that controls the number of dimensions retained in reduced space. As $\epsilon$ decreases, CEBUred is able to recover the reference posterior solution accurately both in mean and credible intervals. We further compare the performance of  CEBUred to that of aBUS-SUS (adaptive Bayesian updating with subset simulation, \cite{Betz2018}), which is a well-established method for nonlinear BIPs in high-dimensions.
\\~\\
The results of our numerical investigations show that CEBUred is a powerful method for solving high-dimensional nonlinear BIPs if the underlying computational model admits a low-dimensional representation, i.e., if the spectrum of $\mathbf{H}$ exhibits fast decay. 
From a computational perspective, CEBUred is particularly useful if the model allows for the cheap evaluation of gradients, e.g., if an adjoint solver is used and the the number of BIP inputs exceeds the number of available observations. In these cases, CEBUred achieves the same accuracy as aBUS-SuS at considerably lower computational expense.
\\~\\
Conversely, if an adjoint solver is not available or not efficient in the sense described above, the gain in computational efficiency provided by dimensionality reduction may be overcompensated by expensive gradient evaluations. In such case, the number of required gradient evaluations could be significantly reduced by using 'data-free likelihood-informed dimension reduction` as recently proposed in \cite{Cui_2021}. There, $\mathbf{H}$ is constructed in expectation over the data such that no knowledge about the posterior density is needed. Consequently, the upper bound between the exact and approximated posterior is controlled in expectation over the data. Following this method, the algorithm of CEBUred could be modified such that $\mathbf{H}$ is constructed at the beginning and therefore, only at this stage model gradient information would be required. Alternatively, one may turn to gradient-free supervised dimension reduction methods in order to identify sutaible subspaces to solve the Bayesian inverse problem.
\section*{Acknowledgments}
We acknowledge support by the German Research Foundation (DFG) through Grants STR 1140/11-1 and PA 2901/1-1.
%

\bibliographystyle{siamplain}
\bibliography{references}

\end{document}